\documentclass{nature}

\usepackage{graphicx}
\usepackage[font=small]{caption}
\usepackage{xcolor}
\usepackage{amsmath} 
\usepackage{booktabs}
\usepackage{float} 

\usepackage{multirow}%
\usepackage{amsmath,amssymb,amsfonts}%
\usepackage{amsthm}%
\usepackage{mathrsfs}%
\usepackage[title]{appendix}%
\usepackage{textcomp}%
\usepackage{manyfoot}%
\usepackage{booktabs}%
\usepackage{algorithm}%
\usepackage{algorithmicx}%
\usepackage{algpseudocode}%
\usepackage{listings}%
\usepackage{braket}
\usepackage{url}
\usepackage{lineno}
\usepackage{amsmath}
\usepackage{tabularx}
\usepackage{makecell}
\usepackage{array}

\title{Enhanced Valley Polarization via Nonlinear Cascaded Quantum-Geometric Selection Rules}

\author
{Quentin Courtade$^{1}$, Sotirios Fragkos$^{1}$, Dominique Descamps$^{1}$, Stéphane Petit$^{1}$, Yann Mairesse$^{1}$, Michael Schüler$^{2,3,\dagger}$, Samuel Beaulieu$^{1,\ddag}$}

\begin{document}

\maketitle

\begin{affiliations}
 \item Universit\'e de Bordeaux - CNRS - CEA, CELIA, UMR5107, F33405 Talence, France
 \item PSI Center for Scientific Computing, Theory and Data, 5232 Villigen PSI, Switzerland 
  \item Department of Physics, University of Fribourg, CH-1700 Fribourg, Switzerland 
  $^{\dagger}$michael.schueler@psi.ch \\
  $^{\ddag}$samuel.beaulieu@u-bordeaux.fr
\end{affiliations}

\begin{abstract}
The quantum geometric properties of Bloch electrons fundamentally govern light–matter interactions and optical selection rules in solids. In semiconducting transition-metal dichalcogenides, circularly polarized excitation near the band edge enables valley-selective interband transitions, providing the basis for valleytronics. While nonlinear optical protocols are being developed to manipulate and probe valley selection rules, they largely rely on band-edge transitions that proceed via virtual intermediate states. Here, we demonstrate a doubly resonant cascaded nonlinear pathway from the valence band to high-lying states, mediated by a real intermediate state whose participation substantially reshapes the valley optical selection rules. Using time- and angle-resolved extreme-ultraviolet photoemission spectroscopy in combination with a time-dependent Lindblad master-equation formalism, we show that this cascaded nonlinear photoexcitation produces a substantially enhanced high-lying valley polarization compared to the conventional linear optical response near the band edge. The extension of the quantum-geometry-based selection rules to the nonlinear regime and high-lying bands offers new perspectives for ultrafast valleytronics and should play a determinant role in strong-field-driven phenomena in quantum materials. 
\end{abstract}

In crystalline solids, the periodic lattice potential shapes the electronic band structure, often giving rise to multiple inequivalent extrema in momentum space, known as valleys. Valleytronics is an emerging field that exploits the valley degree of freedom of electrons as an information carrier, analogous to charge in electronics and spin in spintronics, for the encoding and manipulation of quantum information~\cite{Schaibley16}. Polarization-controlled light-matter interaction offers a particularly appealing way to address the valley degree of freedom, because symmetry-derived optical selection rules enable the selective excitation of individual valleys~\cite{Yao08, Mak12, Zeng12, Cao12, Rodin16, Lin2018, Chen18, Beaulieu2024-qn, Gindl25, pan2025}. This is the case in semiconducting transition metal dichalcogenides (TMDCs), which are widely regarded as prototypical materials for valleytronic applications~\cite{Schaibley16}. In monolayer TMDCs, resonant photoexcitation near the band edge with circularly polarized light enables valley-specific optical transitions at the degenerate K and K$^{\prime}$ valleys, a consequence of the broken inversion symmetry of the hexagonal lattice~\cite{Yao08, Mak12, Zeng12, Cao12, Beaulieu2024-qn}. More specifically, the distinct orbital angular momentum and Berry curvature textures of the electronic bands at the K and K$^{\prime}$ valleys give rise to optical selection rules that govern valley-selective interband transitions. Enhancing valley polarization has long been a central objective in the field, with approaches spanning polarization tailoring~\cite{Sharma23, Tyulnev24, gill2025}, temperature control~\cite{Zhu2025}, pump detuning~\cite{Berghauser2018, Lan24}, strain engineering~\cite{Kumar2025}, and the formation of moiré superlattices~\cite{Dai24, Wu25}. However, the vast majority of experimentally realized valley-polarized states remain confined to energies near the band edge and are typically generated and detected using conventional optical schemes.

Extending beyond the linear regime, nonlinear optical protocols~\cite{Ye14, Xiao2015, Klimmer2021, Herrmann2025-lm, Friedrich2026, tornow2026} have been used to probe excitonic dark states~\cite{Ye14}, reveal local Berry curvature~\cite{tornow2026} and probe optically induced breaking of time-reversal symmetry~\cite{Herrmann2025-lm, Friedrich2026} in TMDCs, for example. While these advances establish a foundation for nonlinear valleytronics, they primarily exploit the well-known spin, orbital, and associated quantum-geometric properties of the electronic states near the band edge (i.e., the top valence band (VB) and the bottom conduction band (CB)). Extending these concepts beyond a two-band picture and simultaneously harnessing the quantum-geometric properties of multiple electronic states could open new avenues for designing light–matter interaction principles in valleytronics and may provide alternative routes for enhancing valley polarization.

It has recently been shown that a specific high-lying conduction band, namely CB+2 in monolayer TMDCs, located at an energy approximately twice that of the band edge, plays a significant role in the optoelectronic properties of TMDCs~\cite{Lin2019, Lin2021, Lin2022} as well as in twisted bilayer TMDCs~\cite{Lin2021t}. The involvement of CB+2 states has been demonstrated to give rise to pronounced quantum interference effects in second-harmonic generation~\cite{Lin2019} and to the formation of high-lying many-body states (excitons and trions) in TMDCs~\cite{Lin2021, Lin2022}. This peculiar excitation pathway involves nonlinear optical transitions or Auger-like processes, characterized by the annihilation of a band-edge exciton to populate high-lying states, leading to upconverted photoluminescence~\cite{Manca2017, Lin2021}. The discovery of this ladder-type three-level system (VB, CB, and CB+2) opens an opportunity to go beyond conventional nonlinear optics near the band edge, which typically relies on in-gap virtual states, by enabling cascaded optical transitions between multiple real bound states. This approach could open a pathway to exploiting the quantum-geometric properties of multiple electronic states for novel, resonantly enhanced multiphoton light–matter interaction mechanisms in valleytronic materials. However, investigating photoexcitation-driven resonant nonlinear valley selection rules, without being constrained by the subset of states accessible through single-photon radiative emission selection rules, requires moving beyond conventional optical detection techniques.

Here, we address this challenge by combining time- and angle-resolved extreme-ultraviolet photoemission spectroscopy (trARPES) with a time-dependent Lindblad master-equation formalism. We investigate the ultrafast nonequilibrium dynamics, multiphoton selection rules, and associated valley polarization resulting from the population of higher-energy conduction bands in 2H-MoTe$_2$. We show that the population of the CB+2 band is short-lived and that the valley polarization of these high-lying states, accessed via cascaded chiroptical resonant excitation, is enhanced relative to that of the CB populated via conventional single-photon transitions near the band edge. We attribute this enhancement to nonlinear valley selection rules arising from the orbital angular momentum and associated quantum-geometrical properties of the multiple bands involved in this doubly resonant multiphoton process.

\begin{figure}[t!]
\centering
\includegraphics[width=0.7\textwidth]{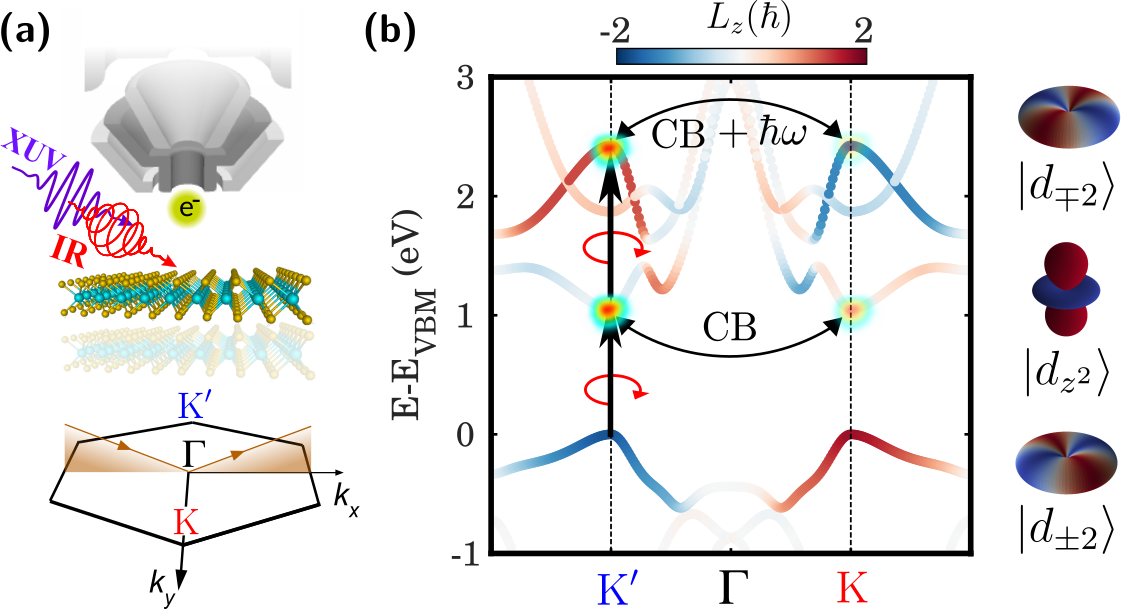}
\caption{\textbf{Scheme of the Experimental Setup and Concept of Enhanced Valley Polarization via Nonlinear Cascaded Quantum-Geometric Selection Rules}. \textbf{(a)} Polarization-tunable infrared pump (1.2~eV, 135~fs, 2.7 mJ/cm$^2$) and XUV (21.6~eV) probe pulses are focused onto a 2H-MoTe$_2$ sample held in front of a time-of-flight momentum microscope, at room temperature and at an incidence angle of 65$^{\circ}$, with the light scattering plane aligned along the crystal mirror plane ($\Gamma$-M direction). \textbf{(b)} Band-resolved local orbital angular momentum ($L_z=m\hbar$) of the W-$d$ orbitals ($\left | d_{m} \right>$, where $m$ is the magnetic quantum number), and associated orbital character of the valence band (VB), first conduction band (CB), and third conduction band (CB+2 or CB+$\hbar\omega$) at K/K$^{\prime}$ valleys, and associated chiroptical multiphoton interband transitions. The curved black arrows represent K$\leftrightarrow$K$^{\prime}$ intervalley scattering.}
\label{Fig1}
\end{figure}

\section{Results} 
Our experimental apparatus~\cite{Fragkos2025-ob} is built around a polarization-tunable ultrafast extreme ultraviolet (XUV) beamline~\cite{Comby22}, coupled to a time-of-flight momentum microscope~\cite{Medjanik17, tkach24, tkach24-2} (see Fig.~\ref{Fig1}(a)). Polarization-tunable infrared (IR) pump pulse (1.2~eV, 135~fs, 2.7~mJ/cm$^2$) and XUV probe pulse (21.6~eV) are focused onto a freshly cleaved 2H-MoTe$_2$ sample held at room temperature, inside the interaction chamber of the momentum microscope end station, allowing the measurement of the nonequilibrium electronic structure and associated occupations of 2H-MoTe$_2$. Further details about the experimental setup can be found elsewhere~\cite{Fragkos2025-ob} and in the Methods. 

Although bulk 2H-MoTe$_2$ is inversion symmetric, each atomic trilayer lacks inversion symmetry. This local symmetry breaking, together with strong spin–orbit coupling, gives rise to a distinctive locking between spin, orbital, valley, and layer degrees of freedom~\cite{Zhang14}. Furthermore, owing to the atomic-scale inelastic mean free path of photoelectrons in XUV photoemission, it has been shown that the spin~\cite{Riley14, Razzoli17, Fanciulli23} and orbital~\cite{Beaulieu20-2, Schuler22-1} textures of the topmost layer can be selectively probed, i.e., features that remain inaccessible in bulk-sensitive measurements. Our experimental approach thus allows us to reveal symmetry-driven selection rules of the monolayer, even if measurements are performed on cleaved bulk crystals~\cite{Riley14, Razzoli17, Fanciulli23, Beaulieu20-2, Schuler22-1}. Moreover, since photoemission provides direct momentum resolution, it does not require an additional control pulse to break time-reversal symmetry via valley-selective optical Stark~\cite{Sie15} and Bloch-Siegert~\cite{Sie17} effects to achieve valley resolution, as is typically required in optical approaches~\cite{tornow2026}. 

To model the valley selection rules, ultrafast dynamics, and photoemission intensities observed in the experiments, we performed density-functional theory (DFT) calculations for monolayer MoTe$_2$ using the \textsc{Quantum Espresso} package~\cite{giannozzi_quantum_2009}, and constructed the associated projective Wannier Hamiltonian using the \textsc{Wannier90} code~\cite{Mostofi08}. The electronic structure and associated band-resolved local orbital angular momentum (OAM, $L_z$) along K-$\Gamma$-K$^{\prime}$ is shown in Fig.~\ref{Fig1}(b). The electronic states at the K/K$^{\prime}$ valleys are dominated by W-$d$ orbitals, which have magnetic quantum numbers $m$ ($\left | d_{m} \right\rangle$), which describe local OAM around the z-axis ($L_z = m\hbar$). While the valence band is characterized by $L_z = \pm 2\hbar$ at K/K$^{\prime}$ valleys ($\left | d_{\pm2} \right>$ orbitals), the associated CB and CB+2 are characterized by $L_z = 0$ ($\left | d_{z^2} \right>$ orbitals) and $L_z = \mp 2\hbar$ ($\left | d_{\mp2} \right>$ orbitals), respectively. This band-resolved local OAM texture, which is intimately related to the local quantum-geometric properties of the Bloch states (i.e., Berry curvature), governs the multiphoton chiroptical selection rules as well as the resulting initial valley polarization (Fig.~\ref{Fig1}(b)). Indeed, as discussed in more detail below, the OAM transferred upon optical transition between two bands determines the helicity preference of the transition. For systems with $C_3$ symmetry, this transfer is given by the difference in OAM between the initial and final bands, evaluated modulo 3 due to the crystal symmetry. More details of the time-dependent Lindblad master equation formalism, as well as the theoretical framework describing light–matter coupling and photoemission processes, are provided in the Methods section and the Supplementary Information.

\begin{figure*}
\begin{center}
\includegraphics[width=\textwidth]{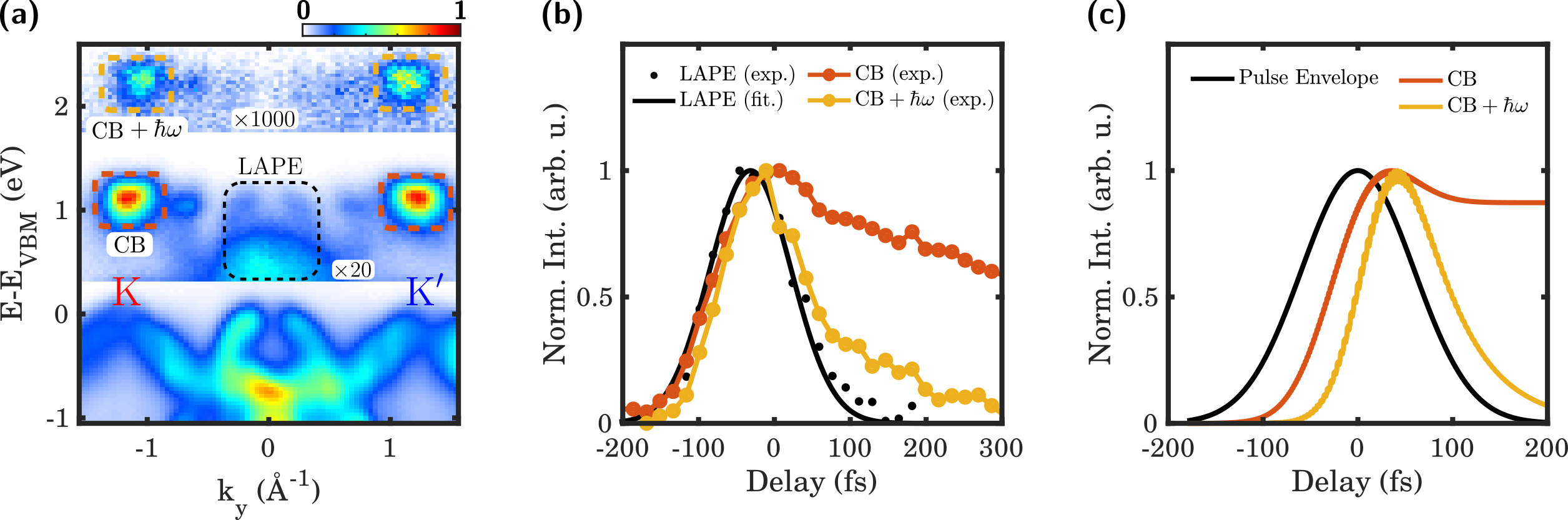}
\caption{\textbf{Nonequilibrium Band Mapping and Ultrafast State-Resolved  Dynamics.}
\textbf{(a)} Normalized energy–momentum cut along the K-$\Gamma$-K$^\prime$ high-symmetry direction recorded at the pump-probe temporal overlap, integrated over all IR pump polarization states (continuous rotation of the IR quarter-wave plate). \textbf{(b)} Experimentally measured time-resolved photoemission intensity three selected energy-momentum windows (shown as colored dashed squares in \textbf{(a)}): i) the replica of the valence band around $\Gamma$ and its Gaussian fit (Laser-Assisted Photoemission - LAPE) in black, ii) the first conduction band (CB) at K and K$^{\prime}$ in orange, and iii) the third conduction band (CB+$\hbar\omega$) at K and K$^{\prime}$ in yellow. \textbf{(c)} Same as \textbf{(b)}, but calculated using the time-dependent Lindblad master-equation formalism.}
\label{fig:nonlin}
\end{center}
\end{figure*}

We begin by investigating light-induced modifications of the band structure occupation in resonantly driven 2H-MoTe$_2$. To this end, we temporally overlapped polarization-tunable 1.2~eV pump pulses with 21.6~eV XUV probe pulses. The polarization state of the pump is continuously modulated by rotating an IR quarter-wave plate, while the energy- and momentum-resolved photoemission intensity is recorded on the fly (while staying at the pump-probe overlap). The corresponding energy-momentum map along the K-$\Gamma$-K$^{\prime}$ high-symmetry direction, integrated over all pump polarization states, is shown in Fig.~\ref{fig:nonlin}(a). First, within the bandgap, some spectral weight around $\Gamma$ (dashed black square) is attributed to a nonresonant two-photon (IR+XUV) process, i.e., laser-assisted photoemission (LAPE) from the valence band. Additionally, excited state populations are observed at the K/K$^{\prime}$ (dashed orange squares) and $\Sigma$ conduction band minima (CBM). In addition, photoemission intensity appears at one pump photon energy above the first CB (dashed yellow squares, $\mathrm{CB} + \hbar\omega$). 

Next, we investigate the ultrafast dynamics of these light-induced spectral features by performing pump-probe measurements using a circularly polarized IR pump. Fig.~\ref{fig:nonlin}(b) shows time-resolved photoemission intensities in selected energy-momentum windows represented by colored dashed squares shown in Fig.~\ref{fig:nonlin}(a). The in-gap LAPE spectral feature around $\Gamma$, delimited by a dashed black square in Fig.~\ref{fig:nonlin}(a), exhibits a characteristic Gaussian-like response ($146 \pm 3$~fs FWHM). This signal represents the optical cross-correlation of our experiment, defining both the absolute timing between the pump and probe pulses and the overall temporal resolution. The spectral feature associated with photoexcited first conduction band states at K and K$^{\prime}$ (CB, in orange) exhibits a population build-up followed by a decay on a picosecond timescale. As recently reported for this material~\cite{girotto25}, the decay of the excited state population can be fitted with a double-exponential function, yielding $\tau_1 = 550 \pm 59~\mathrm{fs}$ and $\tau_2 = 1.4 \pm 0.6~\mathrm{ps}$ (see SI Fig.~S1 for pump-probe measurement over an extended delay range and associated fits). The build-up of transient photoexcited CB states (CB, in orange) is slightly delayed with respect to the cross-correlation (LAPE - in black), as excited-state populations follow the cumulative integral of the pump-pulse envelope. The spectral feature at K and K$^{\prime}$ located at $\mathrm{CB} + \hbar\omega$ (shown in yellow) exhibits an even more pronounced delay with respect to the cross-correlation. This delayed response can be understood as arising from the interplay between population build-up in the first conduction band and subsequent optical excitation from the band edge to higher-lying states. In addition, the spectral feature located at $\mathrm{CB} + \hbar\omega$ is very short-lived. Indeed, applying the same model as for CB to fit the population dynamics yields $\tau_1 = 47 \pm 42~\mathrm{fs}$ and $\tau_2 = 150 \pm 28~\mathrm{fs}$. All these features are qualitatively reproduced by our time-dependent Lindblad master-equation calculations shown in Fig.~\ref{fig:nonlin}(c), with even a better resolution of the delays between the different photoemission signals. Our simulations further reveal that the CB$+\hbar\omega$ signal is strongly suppressed when CB+2 is excluded from the model (or detuned; see Fig.~S2 in the SM), highlighting the resonant nature of this feature. 

\begin{figure*}[t!]
    \centering
    \includegraphics[width=\textwidth]{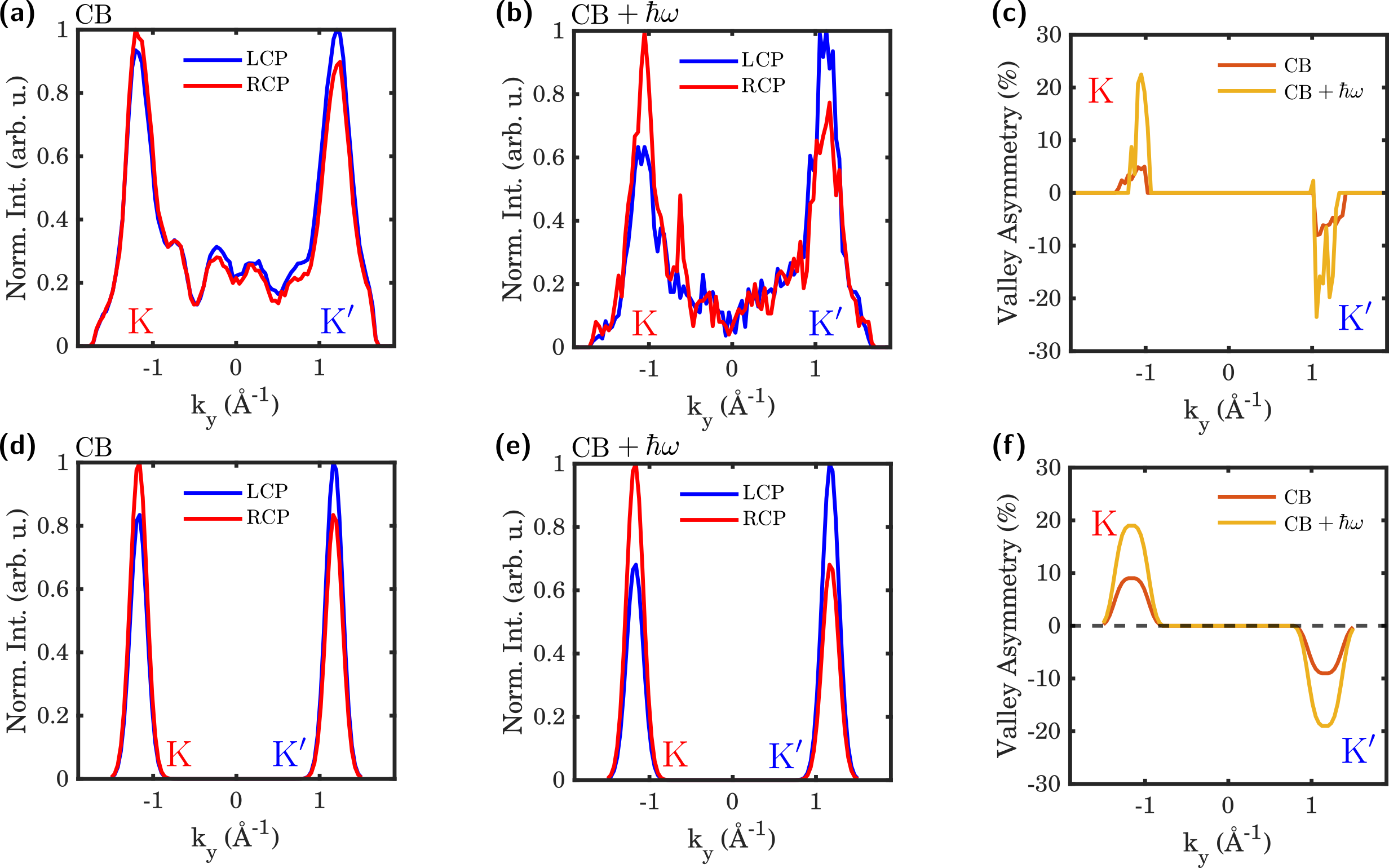}
    \caption{\textbf{Enhanced Valley Polarization in High-Lying States.} \textbf{(a)-(b)} Experimentally measured momentum distribution curves (MDCs) along the K-$\Gamma$-K$^{\prime}$ high-symmetry direction, for LCP (blue) and RCP (red) pump pulses, measured at the pump-probe temporal overlap, at the CB and $\mathrm{CB}+\hbar\omega$ energies, respectively. \textbf{(d)-(e)}  Same as \textbf{(a)-(b)}, but calculated using the time-dependent Lindblad master-equation. \textbf{(c)} and \textbf{(f)} Experimentally measured and calculated valley asymmetry, extracted from the normalized difference between photoemission intensity using LCP pump and RCP pump, at the CB (in orange) and $\mathrm{CB}+\hbar\omega$ (in yellow) energies.}
    \label{fig:valley_pol}
\end{figure*}

After establishing the time-domain signatures associated with the formation of high-lying $\mathrm{CB} + \hbar\omega$ states, we now investigate whether their population can be controlled in a valley-selective manner. Indeed, while resonant single-photon interband transitions near the band edge driven by circularly polarized light are well known to generate valley-polarized carriers, extending these selection rules to higher-lying states via chiroptical multiphoton processes remains an open question and an emerging frontier of valleytronics. Because of their ultrashort-lived nature, our strategy to investigate the valley polarization of high-lying $\mathrm{CB} + \hbar\omega$ states following multiphoton transition is to stay at the pump-probe temporal overlap and extract valley-resolved population for left- (LCP) and right-circularly polarized (RCP) driving pulses.  

In Figs.~\ref{fig:valley_pol}(a) and~(b), we show momentum distribution curves (MDCs) taken at energies corresponding to the first CB and high-lying CB ($\mathrm{CB} + \hbar\omega$) under LCP (blue) and RCP (red) polarized driving fields, along the K-$\Gamma$-K$^{\prime}$ high symmetry direction. The first CB at K/K$^{\prime}$ (Fig.~\ref{fig:valley_pol}(a)) exhibits a small but non-vanishing helicity-dependent valley asymmetry: the K (K$^{\prime}$) valley population is enhanced under RCP (LCP) photoexcitation. At elevated (room) temperature, and under such experimental geometry (see Fig.~S3), such a small valley polarization is not surprising, specifically in 2H-MoTe$_2$, which is known to exhibit smaller valley contrast with respect to other semiconducting TMDCs~\cite{Zhang2022-rt}. Moreover, a small photoemission contribution from the second layer, which would have an opposite valley polarization, could also lead to a reduction of the measured valley asymmetry~\cite{Riley14, Bertoni16}. 

Remarkably, this chiroptical control of valley asymmetry not only survives but is amplified in high-lying $\mathrm{CB} + \hbar\omega$ states (Fig.~\ref{fig:valley_pol}(b)). To quantify the valley asymmetry, we extract the normalized difference of MDCs for RCP and LCP pump, at K and K$^{\prime}$ valleys, for both energy windows (see Fig.~\ref{fig:valley_pol}(c)). The resulting valley asymmetry is substantially ($\sim$ 3 times) larger for the nonlinear multiphoton transition to the high-lying $\mathrm{CB} + \hbar\omega$ states than for the first CB in the linear regime. This behavior, i.e., small valley polarization in CB and strongly enhanced valley polarization at $\mathrm{CB} + \hbar\omega$, is well captured by our time-dependent Lindblad master-equation calculations (Figs.~\ref{fig:valley_pol}(d)-(f)), which also include the contribution of LAPE and ultrafast intervalley scattering.  While some LAPE admixture (from CB) contributes to the measured signal at $\mathrm{CB} + \hbar\omega$, it is helicity- and valley-symmetric at K and K$^{\prime}$~\cite{ValleyFloquet24} and thus cannot account for the measured enhancement of the valley asymmetry. From now on, the $\mathrm{CB} + \hbar\omega$ spectral feature can thus be safely labeled CB+2. A more detailed discussion of the roles of LAPE admixture and ultrafast intervalley scattering can be found in the SI (Figs.~S4 and S5). 

We now turn our attention to the microscopic mechanism underlying this enhanced multiphoton high-lying valley polarization. In the linear-response regime, the connection between dipole transitions and the momentum-space variation of the Bloch wave-function -- the quantum geometry -- is well established~\cite{Yao08, Beaulieu2024-qn, verma25, Yu2025, Li26}. Specifically, the quantum metric (the real part of the quantum geometric tensor - QGT) is involved in governing optical selection rules for linearly polarized light~\cite{Li26}, while the Berry curvature (the imaginary part of the QGT) shapes the selection rules for circularly polarized light. These quantities are intimately tied to the orbital angular momentum of the bands. The quantum geometry of Bloch states has also been shown to play a key role for second- and third-order transitions~\cite{Ahn2022, orenstein_topology_2021, morimoto_topological_2016, lai_third-order_2021}. Specifically, for the two-photon interband transition VB$\rightarrow$CB+2, two excitation channels exist~\cite{aversa_nonlinear_1995}: a resonant cascaded process via the intermediate CB, and a nonresonant, purely geometric contribution governed by the shift vector, which encodes the real-space displacement of the electronic wavepacket during interband optical transitions~\cite{PhysRevB.61.5337}. Owing to the resonant nature of the pump, the cascaded process dominates, and the overall excitation pathway is governed by the corresponding individual dipole selection rules, i.e., VB$\rightarrow$CB and CB$\rightarrow$CB+2. We examine these selection rules directly by inspecting the dipole transition strength $d_{\alpha\rightarrow\alpha^\prime}(\mathrm{K/K'}) = |\hat{\epsilon} \cdot \mathbf{v}_{\alpha\alpha^\prime}(\mathrm{K/K'}) |$, and analyze how it is modulated for transitions between specific bands and across different valleys. The quantitative results of these calculations are presented in Fig.~S3 (Supplementary Information) and are qualitatively summarized in Table~\ref{table:valley}, where we also report the dominant orbital character of the relevant bands. 

\renewcommand{\tabularxcolumn}[1]{>{\centering\arraybackslash}m{#1}}
\renewcommand{\arraystretch}{1}

\begin{table}[t]
\centering
\begin{tabularx}{\linewidth}{!{\vrule width 1.5pt}X|X|X|X!{\vrule width 1.5pt}}
\Xhline{1.5pt}
\textbf{Bands} & \textbf{Orbitals} & $\Delta m_\mathrm{eff}$ at \textbf{K} & $\Delta m_\mathrm{eff}$ at \textbf{K$^{\prime}$} \\
\Xhline{1.5pt}
VB $\rightarrow$ CB & $\left | d_{\pm2} \right> \rightarrow \left | d_{z^2} \right>$ & $+1$ & $-1$ \\
\hline
CB $\rightarrow$ CB+2 & $\left | d_{z^2} \right> \rightarrow \left | d_{\mp2} \right>$ & $+1$ & $-1$ \\
\hline
VB $\rightarrow$ CB+2 & $\left | d_{\pm2} \right> \rightarrow \left | d_{\mp2} \right>$ & $-1$ & $+1$ \\
\Xhline{1.5pt}
\end{tabularx}

\caption{Dipole selection rules for different interband transitions at K and K$^{\prime}$ in terms of the orbital angular momentum transfer $\Delta m$ of the involved $d$ orbitals ($\left | d_{m} \right\rangle$). Due to the $C_3$ symmetry of the system, dipole transitions for right- or left-circularly polarized light are allowed for $\Delta m_\mathrm{eff} = \Delta m\, \mathrm{mod}\,3 = \pm 1$.}

\label{table:valley}
\end{table}

The selection rules for the cascaded two-photon transitions can be rationalized by inspecting the angular momentum transfer $\Delta m$. In periodic solids, the OAM is not conserved as in atoms, as the lattice can absorb or provide additional OAM. To this end, we inspect the OAM of the relevant bands and take into account that OAM transfer is defined modulo 3 for $C_3$ symmetry~\cite{PhysRevLett.115.115502}, see Table~\ref{table:valley}. For the VB$\rightarrow$CB transition at, e.g. K valley ($\left| d_{+2}  \right> \rightarrow \left| d_{0} \right>$), we find $\Delta m = -2 \equiv +1$. Similarly, the CB$\rightarrow$CB+2 transition ($ \left| d_{0} \right> \rightarrow \left| d_{-2} \right>$) requires the OAM transfer $\Delta m = -2 \equiv + 1$. Hence, at a given valley, both transitions are allowed for the same helicity of the circularly polarized photons. These selection rules are reversed for a single-photon transition between VB ($\left| d_{\pm2} \right>$) and CB+2 ($\left| d_{\mp2} \right>$), in good agreement with the pioneering work of Lin \textit{et al.}~\cite{Lin2022}. Because the transitions from VB to CB and from CB to CB$+2$ share the same helicity preference, the valley chiroptical selection rules are effectively applied twice within this two-photon excitation pathway. This leads to an enhanced valley polarization in CB$+2$ compared to CB. 

\section{Discussion}
Our results show that while the population within the first conduction band at the K/K$^{\prime}$ valley is characterized by a relatively long picosecond lifetime, as reported elsewhere~\cite{girotto25}, and moderate valley polarization, the population of high-lying CB+2 states is very short-lived and displays a significantly enhanced valley polarization. We attribute this valley polarization enhancement to nonlinear selection rules that emerge from the unique orbital angular momentum texture and associated quantum-geometric nature of the bands involved in this ladder-type three-level system. This finding extends recently demonstrated nonlinear chiral valley selection rules at the band edge~\cite{Cheng19, Herrmann2025-lm} through virtual states, to the doubly resonant regime involving real electronic states, thereby enabling nonlinear access to enhanced valley-polarization in high-lying bands. 

More broadly, our work can be viewed as an important step toward bridging the gap between the well-understood linear light-matter coupling phenomena underlying valleytronics near the band edge and the more complex nonlinear phenomena emerging in strong-field-driven solids. Indeed, the strong-field physics community is advancing new approaches to drive coherent electronic responses in solids using ultrafast tailored lightwaves~\cite{Gucci2026} and topological optical fields~\cite{Tyulnev24}, as well as to probe light-driven band structures~\cite{Uzan22} and interband Berry phases~\cite{Uzan24} in laser-driven crystals. Our results establish a framework for extending quantum-geometry-based selection rules to the nonlinear regime and high-lying bands, whose role is essential in strong-field-driven quantum materials.

\section{Methods}
\subsection{Experimental setup}\label{sec:exp}
The experimental setup is articulated around a polarization-tunable ultrafast extreme ultraviolet (XUV) beamline~\cite{Comby22} coupled to a momentum microscope end-station~\cite{tkach24, tkach24-2, Fragkos2025-ob}. Our beamline is pumped by a high-repetition-rate Yb fiber laser (166~kHz, 1030~nm, 135~fs, 50~W, Amplitude Laser Group). For the HHG-based XUV probe arm, a significant portion of the laser beam is frequency-doubled in a BBO crystal to generate a few watts of 515~nm pulses. After beam size adjustment and spatial profile tailoring to create an annular beam, these pulses are focused into an argon gas jet to produce XUV radiation through high-order harmonic generation (HHG). To reject the 515~nm driving beam from the XUV beamlet, we use sequential spatial filtering with pinholes. The 9th harmonic of the 515~nm driver (21.6~eV) is spectrally isolated via reflections onto Sc/SiC multilayer XUV mirrors (NTTAT) and propagation through a 200~nm thick Sn filter (Luxel). For the IR pump arm, a small fraction of the fundamental laser pulse (1030~nm, 135~fs) is utilized. The IR pump and XUV probe pulses are collinearly recombined using a drilled mirror. They are subsequently focused onto the sample, achieving typical spot sizes of 70 $\mu$m $\times$ 140 $\mu$m for the IR pump and 45 $\mu$m $\times$ 35 $\mu$m for the XUV probe, respectively. Concerning sample preparation, bulk 2H-MoTe$_2$ samples (HQ Graphene) are glued on the sample holder (Flag style - Ferrovac), and a cylindrical ceramic post (Umicore) is glued on the top surface of the sample using UHV-compatible conductive epoxy (Epo-tek). The samples are then cleaved by striking the ceramic post in an ultrahigh-vacuum environment at a base pressure of 1$\times$10$^{-10}$ mbar. The freshly cleaved sample is then introduced into a motorized hexapod for precise sample alignment within the main chamber (base pressure: 2$\times$10$^{-10}$ mbar, room temperature). Time-resolved photoemission measurements are performed using a custom time-of-flight momentum microscope equipped with an advanced front lens offering multiple operational modes (GST mbH)~\cite{tkach24, tkach24-2}. For data post-processing, we employ an open-source data workflow~\cite{Xian20, Xian19_2} to efficiently transform raw single-event datasets into calibrated, binned hypervolumes of the desired dimensions. More details about the experimental setup can be found in Ref.~\cite{Fragkos2025-ob}.

\section{Details on the calculations}
To model the dynamics and photoemission intensities observed in the experiments, we performed density-functional theory (DFT) calculations for a monolayer of MoTe$_2$. Due to the short mean-free path of the photoelectrons, the time-resolved ARPES (trARPES) signal is dominated by the first layer of the bulk sample; moreover, the layer-resolved selection rules for dipole transitions are well described by individual monolayers in the vicinity of the K/K' points. We performed the DFT calculations for monolayer MoTe$_2$  using the \textsc{Quantum Espresso} package (plane-wave cutoff: 80~Ry, functional: PBE), and in the second step we constructed the projective Wannier Hamiltonian using the \textsc{Wannier90} code. 

Using the band energies $\varepsilon_\alpha(\mathbf{k})$ and velocity matrix elements $\mathbf{v}_{\alpha\alpha^\prime}(\mathbf{k})$, we built an effective Hamiltonian for the relevant bands at K and K':
\begin{align}
    \label{eq:heff}
    H_{\mu\mu^\prime}(t) = \varepsilon_\mu \delta_{\mu \mu^\prime} - \mathbf{A}(t)\cdot \mathbf{v}_{\mu\mu^\prime} \ .
\end{align}
Here, the composite index $\mu = 1,\dots,2 n_B$ labels both the band $\alpha = 1,\dots,n_B$ and the valley K / K' degree of freedom. We include $n_B=4$ bands in each valley: the top valence band and the first three conduction bands. To compensate for the difference in the band structures of bulk and monolayer, and to account for the underestimated band gap, we adjusted the energies $\varepsilon_\mu$, increasing the band gap by 75~meV to match the experimentally observed gap. The light-matter coupling is introduced in Eq.~\eqref{eq:heff} through the vector potential of the pump $\mathbf{A}(t)$.

To incorporate the intervalley scattering, we combine the time-dependent Hamiltonian~\eqref{eq:heff} with the Lindblad master equation:
\begin{align}
    \label{eq:lindblad_eom}
    \frac{d}{dt} \boldsymbol{\rho}(t) = - i [\mathbf{H}(t), \boldsymbol{\rho}(t)] + \sum_s \frac{1}{\tau_s}\boldsymbol{\mathcal{L}}_s \boldsymbol{\rho}(t) \ .
\end{align}
Note that we used compact matrix notation here. In Eq.~\eqref{eq:lindblad_eom}, $s$ labels all possible scattering mechanisms, each described by a Lindbladian $\boldsymbol{\mathcal{L}}_s$. We consider the following scattering processes: (i) intervalley scattering between the bottom conduction bands $\mu = 2 \leftrightarrow \mu = 6$ (time constant $\tau_A$), (ii) intervalley scattering scattering between the highest conduction bands $\mu = 4 \leftrightarrow \mu = 8$ (time constant $\tau_B$), and (iii) pure dissipative decay of the highest conduction bands (time constant $\tau_C$). Choosing $\tau_A=20$~fs and $\tau_C=50$~fs, the simulated population dynamics reproduces the intervalley scattering CB$\leftrightarrow$CB and the decay of the CB+2 signal well. The intervalley scattering time is fixed at $\tau_B=40$~fs for the results presented in the main text. The lower scattering rate compared to CB is consistent with the reduced phase space for scattering events due to the steeper dispersion of CB+2. In the Supplementary Information, we investigated the effect of varying $\tau_B$. In all scenarios, the valley polarization of CB+2 is significantly enhanced compared to CB.

The Lindblad master equation~\eqref{eq:lindblad_eom} is solved assuming an electric field pulse
\begin{align}
    \label{eq:pump_field}
    \mathbf{E}(t) = F(t)\mathrm{Re}[\hat{\epsilon}e^{-i \omega_\mathrm{pump} t}] \ .
\end{align}
For the envelope function, we chose a Gaussian with a peak field strength of $E_0=1.6$~MV/cm and FWHM of 140~fs. The polarization vector $\hat{\epsilon}$ was determined as in Ref.~\cite{ValleyFloquet24} as the transmitted field obtained from Fresnel's equations. 

\subsection{Multiphoton Valley Selection Rules}
Additional insights can be gained from expressing the two-photon transition in terms of Fermi's Golden rule. Employing second-order time-dependent perturbation theory, we obtain
\begin{align}
    \label{eq:two_photon_fermi}
    P_\nu(\omega) \propto|M^{(2)}_{\nu\mu}|^2 \delta(\varepsilon_\nu - \varepsilon_\mu - 2\omega_\mathrm{pump})
\end{align}
for the population in the conduction band $\nu$. The two-photon matrix element is defined by
\begin{align}
\label{eq:two_photon_mel}
    M^{(2)}_{\nu\mu} = \sum_{\mu^\prime} \frac{(\hat{\epsilon} \cdot\mathbf{A}_{\nu\mu^\prime})(\hat{\epsilon} \cdot\mathbf{A}_{\mu^\prime\mu})}{\varepsilon_{\mu^\prime} - \varepsilon_\mu - \omega_\mathrm{pump}} \ ,
\end{align}
where $\mu$ labels the valence band, $\mu^\prime$ are intermediate states (here CB), and $\nu$ stands for CB+2. We have expressed the matrix elements in Eq.~\eqref{eq:two_photon_mel} in dipole gauge here; $\mathbf{A}_{\mu\mu^\prime}$ thus denotes the Berry connection. Following Aversa and Sipe~\cite{aversa_nonlinear_1995}, the sum over all states $\mu^\prime$ in Eq.~\eqref{eq:two_photon_mel} can be split into a resonant cascade-type term and a nonresonant purely geometric term:
\begin{align}
    \label{eq:two_photon_mel_split}
    M^{(2)}_{\nu\mu} &= \sum_{\mu^\prime\ne \nu, \mu} \frac{(\hat{\epsilon} \cdot\mathbf{A}_{\nu\mu^\prime})(\hat{\epsilon} \cdot\mathbf{A}_{\mu^\prime\mu})}{\varepsilon_{\mu^\prime} - \varepsilon_\mu - \omega_\mathrm{pump}} \nonumber \\
    &\quad + \frac{1}{\omega_\mathrm{pump}} \hat{\epsilon} \cdot D_{\mathbf{k}}[\hat{\epsilon}\cdot \mathbf{A}_{\nu\mu}] \ .
\end{align}
Here, $D_{\mathbf{k}}$ denotes the covariant derivative,
\begin{align*}
    D_{\mathbf{k}} O_{\mu\mu^\prime}(\mathbf{k}) = \nabla_{\mathbf{k}}O_{\mu\mu^\prime}(\mathbf{k}) - i O_{\mu\mu^\prime}(\mathbf{k}) (\mathbf{A}_{\mu\mu}(\mathbf{k}) - \mathbf{A}_{\mu^\prime\mu^\prime}(\mathbf{k})) \ .
\end{align*}
Thus, the second term in Eq.~\eqref{eq:two_photon_mel_split} encodes higher-order quantum geometric effects, such as the shift vector. Importantly, it plays an important role for off-resonant two-photon processes. Under the resonant condition as in the experiments, the cascade process (first term in Eq.~\eqref{eq:two_photon_mel_split}) dominates and determines the valley selection rules.

The selection rules for the resonant cascade two-photon transitions can be rationalized by inspecting the angular momentum transfer $\Delta m$. To this end, we inspect the orbital angular momentum (OAM) of the relevant bands and take into account that OAM transfer is defined modulo 3 for $C_3$ symmetry. For the VB$\rightarrow$CB transition at K ($d_{+2} \rightarrow d_{0}$), we find $\Delta m = -2 \equiv +1$. Similarly, the CB$\rightarrow$CB+2 transition ($d_{0} \rightarrow d_{-2}$) requires the OAM transfer $\Delta m = -2 \equiv + 1$. Hence, both transitions are driven by the same helicity of the circularly polarized photons.

\subsection{Simulation of trARPES}
In order to directly compare the simulation with the experiments, we computed the trARPES intensity. Employing a simplified version of the formalism from Ref.~\cite{schuler_theory_2021-1}, we define the trARPES intensity as
\begin{align}
\label{eq:trarpes}
    I(\omega, \tau) = \mathrm{Im}\int^\infty_{0}dt \int^t_{0}dt'\, & s(t,\tau) s(t', \tau) \mathrm{Tr}[G^<(t',t)] \nonumber \\ &\quad \times e^{i[ \omega(t-t') -   \phi_V(t,t')]} \ .
\end{align}
Here, the lesser Green's function is reconstructed from the generalized Kadanoff-Baym ansatz (see Ref.~\cite{schuler_theory_2021-1}), while $s(t,\tau)$ denotes the envelope of the probe pulse centered at time delay $\tau$. The effect of laser-assisted photoemission (LAPE) is captured by the Volkov phase
\begin{align}
    \phi_V(t,t') = - \int^t_{t'}dt''\, \mathbf{A}(t'') \cdot \mathbf{p} \ ,
\end{align}
where $\mathbf{A}(t)$ is the vector potential corresponding to the pump field~\eqref{eq:pump_field}, while $\mathbf{p}$ is the momentum of the photoelectron. The in-plane component is fixed to either K or K' due to in-plane momentum conservation of the photoemission process; the out-of-plane component is determined from energy conservation, yielding $p_\perp \approx 0.9$ bohr. To simplify the calculation, we did not compute the trARPES intensity~\eqref{eq:trarpes} as a function of $\mathbf{k}$, but only at $\mathbf{k}=$K, K', and described the momentum dependence with a Gaussian with a broadening matching the experimental results. The time delay $\tau$ is chosen, consistent with the experiments, to maximize the intensity. In the SI, using this approach, we performed some simulations to investigate the role of LAPE in the experiments. 

\section*{Data Availability}
The data that support the findings of this article will be openly available on Zenodo upon publication. 

\section*{Acknowledgements}
We thank Nikita Fedorov, Romain Delos, Pierre Hericourt, Rodrigue Bouillaud, Laurent Merzeau, and Frank Blais for technical assistance.  We thank Baptiste Fabre for implementing and maintaining the data binning code. We acknowledge the financial support of the IdEx University of Bordeaux/Grand Research Program ``GPR LIGHT". This work is part of the ULTRAFAST and TORNADO projects of PEPR LUMA and was supported by the French National Research Agency, as a part of the France 2030 program, under grants ANR-23-EXLU-0002 and ANR-23-EXLU-0004. We acknowledge support from ERC Starting Grant ERC-2022-STG No.101076639, Quantum Matter Bordeaux, AAP CNRS Tremplin, and AAP SMR from Université de Bordeaux. S.F. acknowledges funding from the European Union’s Horizon Europe research and innovation programme under the Marie Skłodowska-Curie 2024 Postdoctoral Fellowship No 101198277 (TopQMat). Q.C. acknowledges funding from the TERAQUANTUM project of the Région Nouvelle-Aquitaine. Funded by the European Union. This research was supported by the
NCCR MARVEL, a National Centre of Competence in Research, funded by the Swiss National Science Foundation (grant number 205602). Views and opinions expressed are however those of the author(s) only and do not necessarily reflect those of the European Union. Neither the European Union nor the granting authority can be held responsible for them. 

\section*{Author contributions}
S.B. and M.S. conceived the research project. Q.C., S.F., Y.M., and S.B. performed the experiments. Q.C. and S.B. analyzed the experimental data. D.D. and S.P. participated in maintaining the laser system. M.S. developed the theoretical framework and analyzed the results from the simulations. S.B. and M.S. wrote the first draft of the manuscript. Q.C, S.F., D.D., and Y.M. participated in commenting and revising the manuscript. 

\clearpage

\setcounter{figure}{0} 
\renewcommand{\figurename}{Fig.}
\renewcommand{\thefigure}{S\arabic{figure}}

\setcounter{section}{0} 
\renewcommand{\thesection}{S\arabic{section}}

\setcounter{subsection}{0} 
\renewcommand{\thesubsection}{S\arabic{section}.\arabic{subsection}}

\setcounter{equation}{0} 
\renewcommand{\theequation}{S\arabic{equation}}

\setcounter{table}{0}  
\renewcommand{\thetable}{S\arabic{table}}

\begin{center}
    {\Large \textbf{Supplementary information for: \\ Enhanced Valley Polarization via Nonlinear Cascaded Quantum-Geometric Selection Rules}}
\end{center}

\section{Population dynamics of CB and CB+$\hbar\omega$}

Using pump–probe measurements over an extended delay range (up to 5~ps), we extract the population lifetimes of excited carriers in the CB and CB+$\hbar\omega$. As recently reported for this material~\cite{girotto25}, the decay of the excited-state population can be described by a double-exponential function multiplied by a cumulative distribution function to account for the population build-up. Applying this fitting procedure to carriers in the first CB yields $\tau_1 = 550 \pm 59~\mathrm{fs}$ and $\tau_2 = 1.4 \pm 0.6~\mathrm{ps}$. Using the same model to fit the population dynamics in CB+$\hbar\omega$ gives $\tau_1 = 47 \pm 42~\mathrm{fs}$ and $\tau_2 = 150 \pm 28~\mathrm{fs}$.

\begin{figure}[H]
\begin{center}
\includegraphics[width=0.7\textwidth]{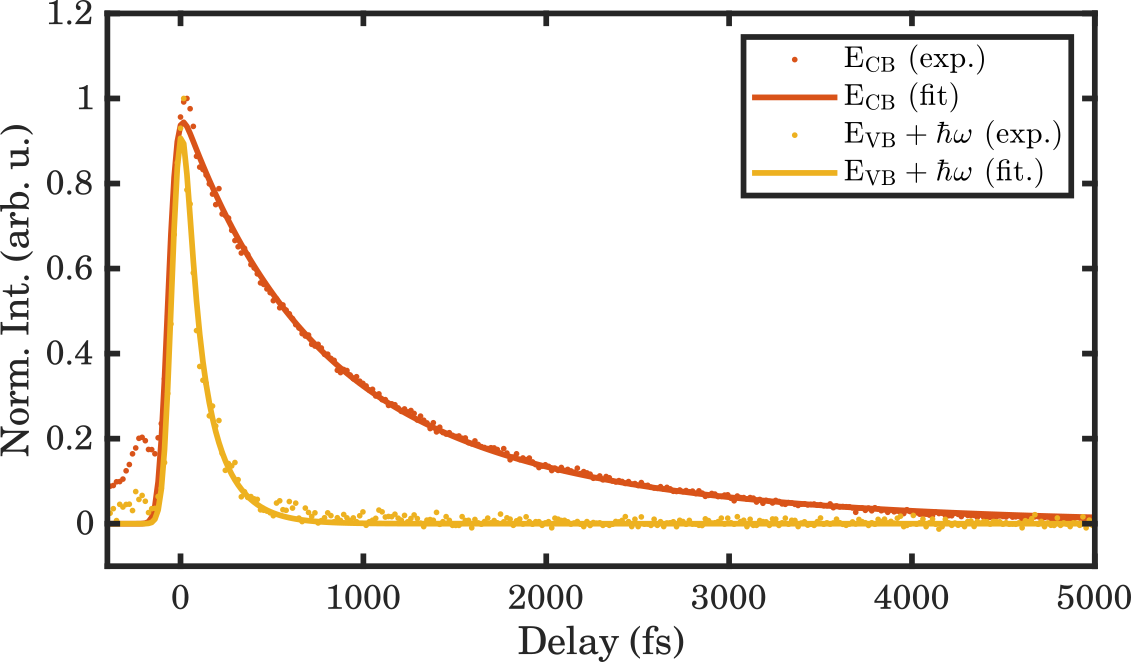}
\caption{Normalized momentum-integrated population dynamics for CB (orange dots) and CB+$\hbar\omega$ (yellow dots) and associated double exponential decays fits (orange and yellow lines). \label{fig:delays}}
\end{center}
\end{figure}

Note that we use the nomenclature CB+$\hbar\omega$ and CB+2 somewhat interchangeably in the following. In the manuscript, we label the measured photoemission signal as CB+$\hbar\omega$ until we conclude that this signal predominantly arises from population in CB+2, rather than from LAPE of the CB.

\section{Role of detuning}

\begin{figure}[H]
\centering
\includegraphics[width=0.85\textwidth]{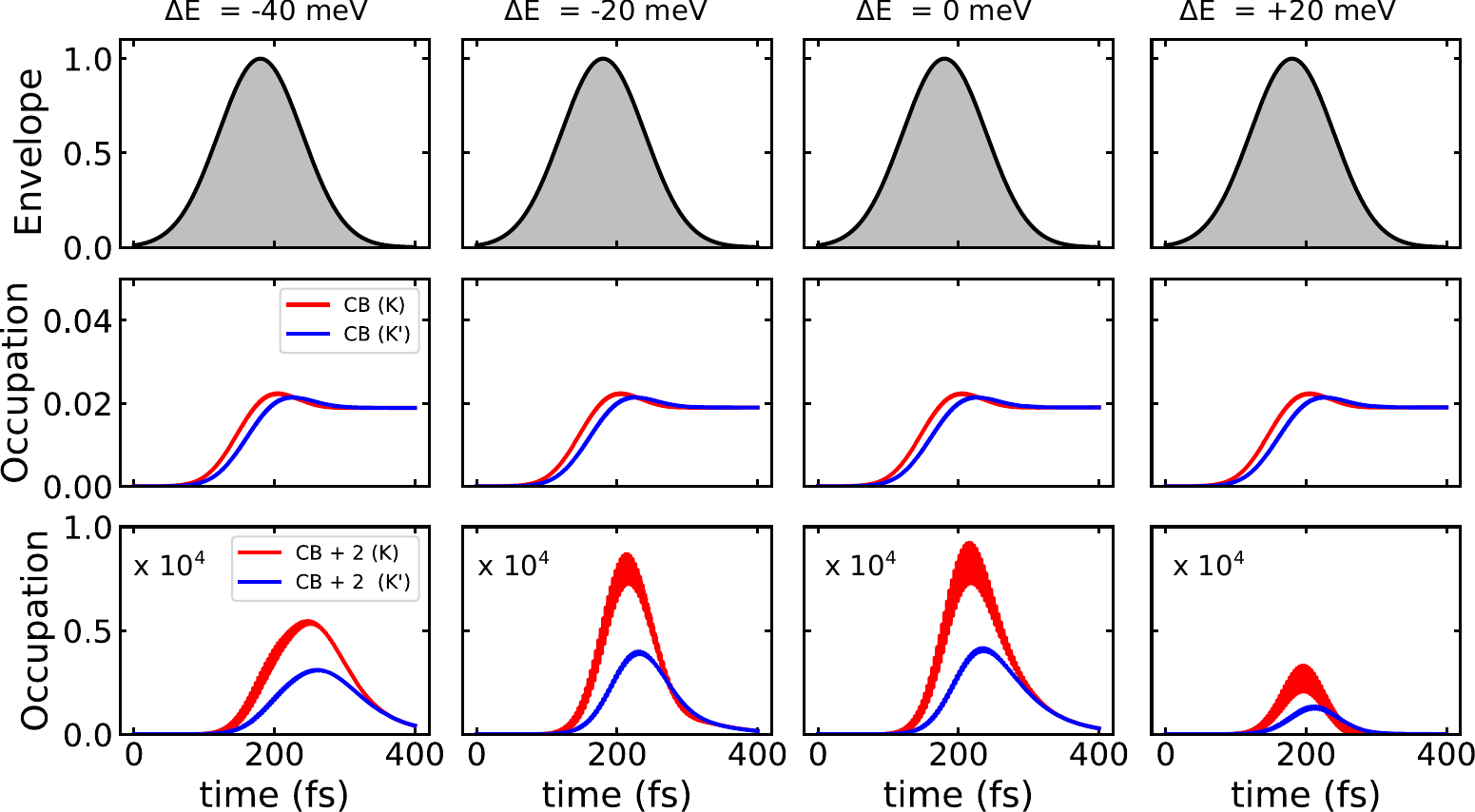}
\caption{Population dynamics for different energy shifts $\Delta E$ of the highest conduction band (CB+2). The occupation of CB+2 has been multiplied by $10^4$ for better visibility. \label{fig:detuning}}
\end{figure}

To investigate the effects of detuning, we introduce the energy shift $\Delta E$ of the highest conduction bands relative to the peak position in the experiments. We then solve the Lindblad master equation (Eq. 2 in the Methods) for every value of $\Delta E$ and extract the populations. The result is presented in Fig.~\ref{fig:detuning} for the values of $\tau_{A,B,C}$ as in the main text. Varying $\Delta E$, we observe a decrease in the population in CB+2. The effect is asymmetric around $\Delta E = 0$ due to the specific scattering model. Nevertheless, the suppression of the CB+2 population upon detuning the position of the higher conduction band demonstrates the importance of the resonant enhancement of both the real and virtual populations. 

\section{Valley selection rules}

\begin{figure}[H]
\centering
\includegraphics[width=0.7\textwidth]{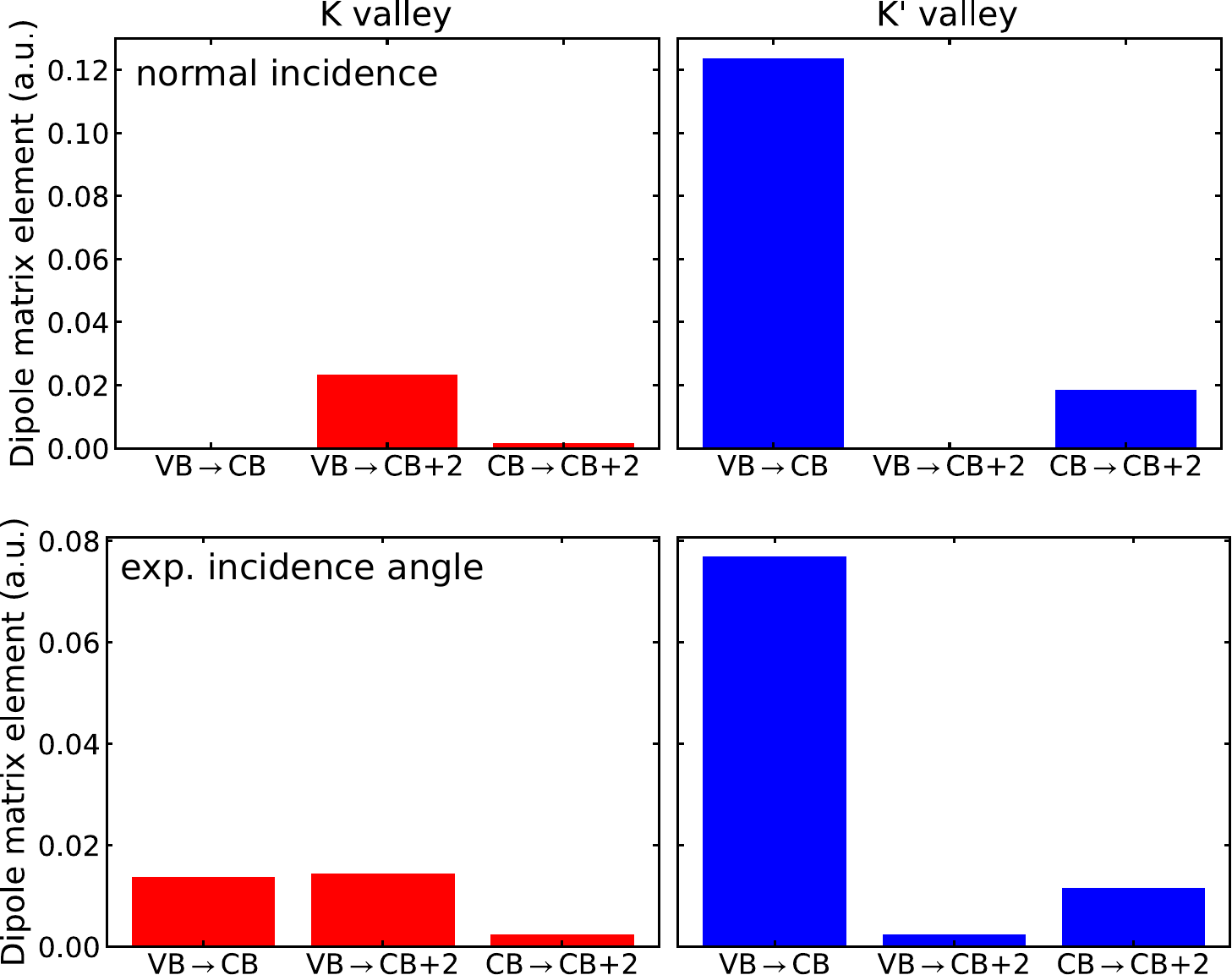}
\caption{Dipole transition strength for right-handed circularly polarized pump impinging in normal incidence (top panels) and under experimental geometry, i.e., at an angle of incidence of 65$^{\circ}$ (bottom panels). \label{fig:dipole_sel} }
\end{figure}

To better understand the valley polarization in the CB and higher CBs, we investigate the selection rules, encoded in the velocity matrix elements, directly. We define the dipole transition strength $d_{\alpha\rightarrow\alpha^\prime}(\mathrm{K/K'}) = |\hat{\epsilon} \cdot \mathbf{v}_{\alpha\alpha^\prime}(\mathrm{K/K'}) |$, which is depicted in Fig.~\ref{fig:dipole_sel}. 

At normal incidence, the effective polarization of the transmitted field remains purely circularly polarized. As a result, the transition VB$\rightarrow$CB is almost 100\% valley selective (note that electron-phonon scattering is absent here). Similarly, the transition from CB to the highest CB (CB+2) is strongly valley-selective. The direction transition VB$\rightarrow$CB+2 is possible in the opposite valley; this process is significantly weaker than the VB$\rightarrow$CB$\rightarrow$CB+2 pathway.
Considering the oblique incidence as in the experiments (angle of incidence of 65$^{\circ}$), the picture qualitatively remains the same. The only difference is a slightly reduced valley polarization.

\section{Role of Laser-Assisted Photoemission}

\begin{figure}[H]
\centering
\includegraphics[width=0.6\textwidth]{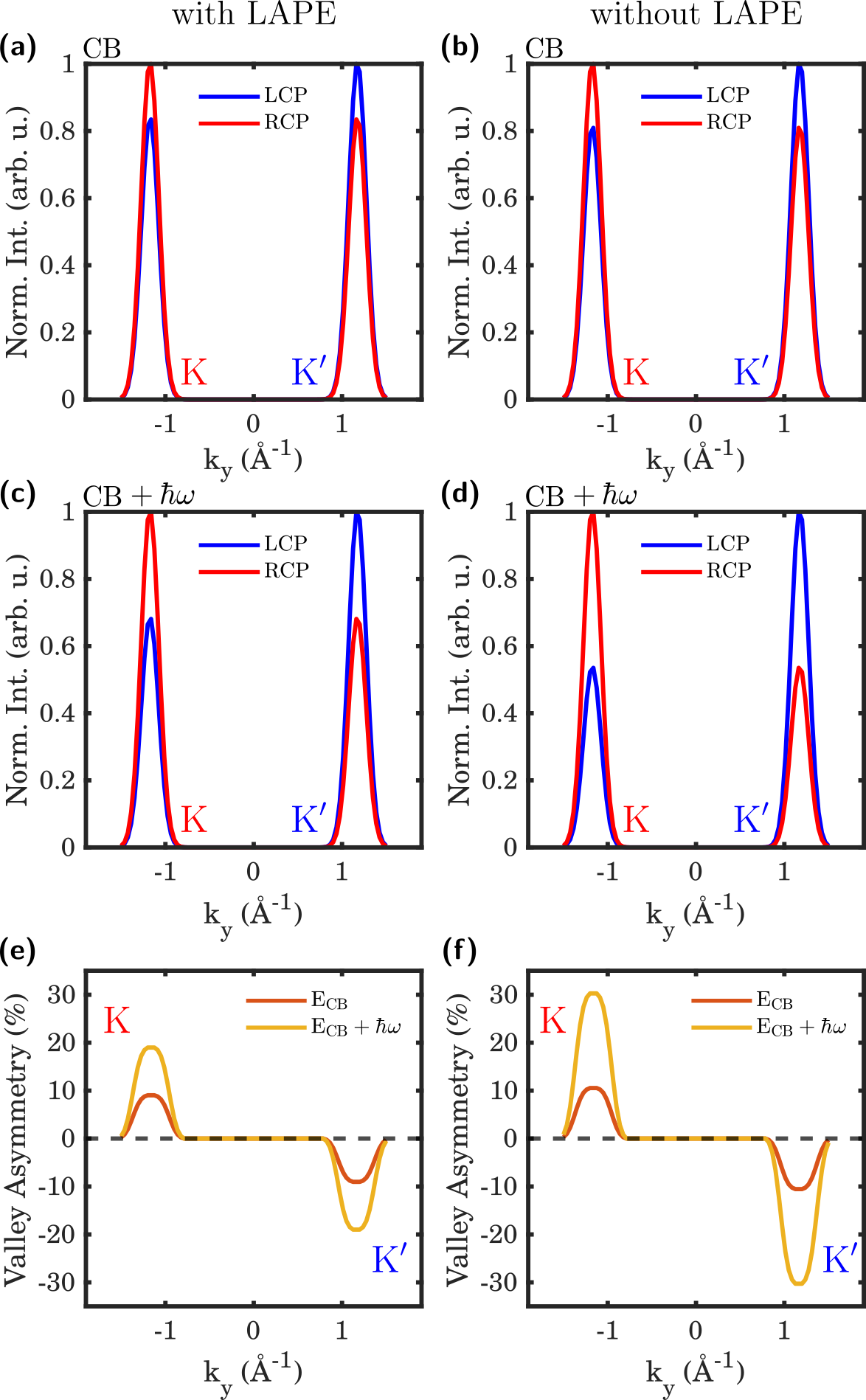}
\caption{\textbf{(a)} and \textbf{(c)} Time-dependent Lindblad master-equation calculations, including the effect of LAPE, of the momentum distribution curves (MDCs) along K-$\Gamma$-K$^{\prime}$ high-symmetry direction, for LCP (blue) and RCP (red) pump pulses, at the CB and $\mathrm{CB}+\hbar\omega$ energies, respectively. \textbf{(e)} Associated valley asymmetry, extracted from the normalized difference between photoemission intensity using LCP and RCP, at the CB (in orange) and $\mathrm{CB}+\hbar\omega$ (in yellow) energies. \textbf{(b)}, \textbf{(d)} and \textbf{(f)} Same as \textbf{(a)}, \textbf{(c)} and \textbf{(e)}, but without including the contribution of LAPE. 
\label{fig:MDCLAPE} }
\end{figure}

Excluding the Volkov phase (see Methods) allows us to switch off the LAPE process and study its effects (Fig.~\ref{fig:MDCLAPE}). Indeed, due to the resonant nature of the transition VB$\rightarrow$CB, LAPE has very little influence on the observed valley polarization in the CB (Figs.~\ref{fig:MDCLAPE}(a) and~(b)). This changes for CB+2 (Figs.~\ref{fig:MDCLAPE}(c) and~(d)): the valley polarization of the trARPES signal at CB + $\hbar\omega$ is reduced by including LAPE. This result is explained by the fact that the LAPE contribution is independent on the helicity of the pump, thus adding an unpolarized background to the intensity stemming from the population of CB+2. Nevertheless, the enhancement of the valley polarization of the CB$+\hbar\omega$ feature compared to CB is qualitatively robust even in the presence of LAPE. 

\section{Role of ultrafast intervalley scattering}

\begin{figure}[H]
\centering
\includegraphics[width=0.6\textwidth]{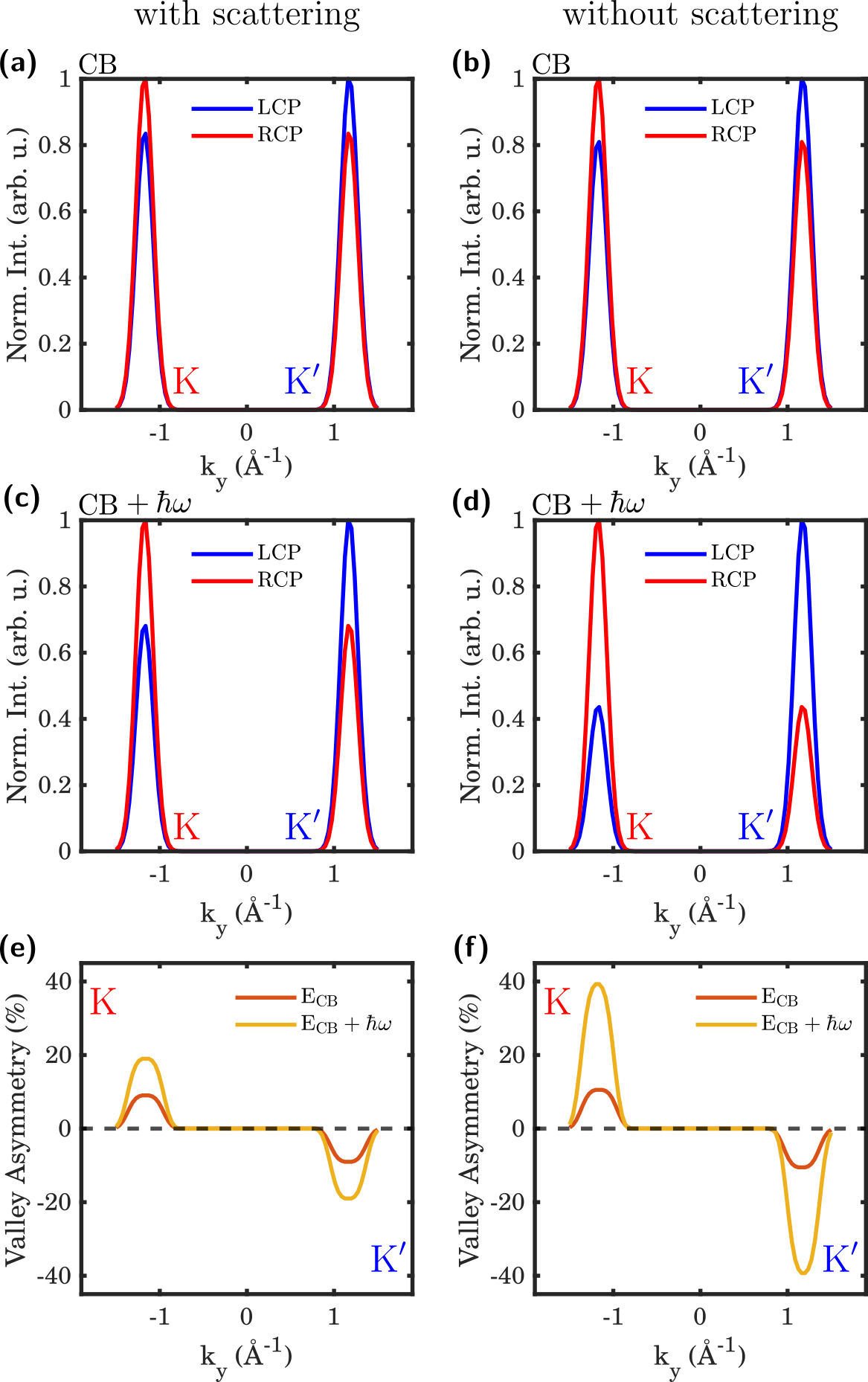}
\caption{\textbf{(a)} and \textbf{(c)} Time-dependent Lindblad master-equation calculations, including the effect of ultrafast intervalley scattering, of the MDCs along K-$\Gamma$-K$^{\prime}$ high-symmetry direction, for LCP (blue) and RCP (red) pump pulses, at the CB and $\mathrm{CB}+\hbar\omega$ energies, respectively. \textbf{(e)} Associated valley asymmetry, extracted from the normalized difference between photoemission intensity using LCP and RCP, at the CB (in orange) and $\mathrm{CB}+\hbar\omega$ (in yellow) energies. \textbf{(b)}, \textbf{(d)} and \textbf{(f)} Same as \textbf{(a)}, \textbf{(c)} and \textbf{(e)}, but without including the contribution of ultrafast intervalley scattering for the higher conduction bands. 
\label{fig:MDCscatt} }
\end{figure}

To capture the role of intervalley scattering processes in the higher conduction bands, we performed simulations with various choices for the intervalley scattering time $\tau_B$. In Fig.~\ref{fig:MDCscatt} we compare the photoemission intensity for the case of $\tau_B = 40$~fs (left panels) and for the case of $\tau_B=\infty$ (right-hand side panels). The ultrafast intervalley scattering among the CB+2 states reduces the observed valley polarization (Figs.~\ref{fig:MDCscatt}(c) and~(d)), but it can not suppress it.

\clearpage


\begin{thebibliography}{10}
\expandafter\ifx\csname url\endcsname\relax
  \def\url#1{\texttt{#1}}\fi
\expandafter\ifx\csname urlprefix\endcsname\relax\def\urlprefix{URL }\fi
\providecommand{\bibinfo}[2]{#2}
\providecommand{\eprint}[2][]{\url{#2}}

\bibitem{Schaibley16}
\bibinfo{author}{Schaibley, J.~R.} \emph{et~al.}
\newblock Valleytronics in 2D materials.
\newblock \emph{\bibinfo{journal}{Nature Reviews Materials}} \textbf{\bibinfo{volume}{1}}, \bibinfo{pages}{16055} (\bibinfo{year}{2016}).
\newblock \urlprefix\url{https://doi.org/10.1038/natrevmats.2016.55}.

\bibitem{Yao08}
\bibinfo{author}{Yao, W.}, \bibinfo{author}{Xiao, D.} \& \bibinfo{author}{Niu, Q.}
\newblock Valley-dependent optoelectronics from inversion symmetry breaking.
\newblock \emph{\bibinfo{journal}{Phys. Rev. B}} \textbf{\bibinfo{volume}{77}}, \bibinfo{pages}{235406} (\bibinfo{year}{2008}).
\newblock \urlprefix\url{https://link.aps.org/doi/10.1103/PhysRevB.77.235406}.

\bibitem{Mak12}
\bibinfo{author}{Mak, K.~F.}, \bibinfo{author}{He, K.}, \bibinfo{author}{Shan, J.} \& \bibinfo{author}{Heinz, T.~F.}
\newblock Control of valley polarization in monolayer MoS$_2$ by optical helicity.
\newblock \emph{\bibinfo{journal}{Nature Nanotechnology}} \textbf{\bibinfo{volume}{7}}, \bibinfo{pages}{494--498} (\bibinfo{year}{2012}).
\newblock \urlprefix\url{https://doi.org/10.1038/nnano.2012.96}.

\bibitem{Zeng12}
\bibinfo{author}{Zeng, H.}, \bibinfo{author}{Dai, J.}, \bibinfo{author}{Yao, W.}, \bibinfo{author}{Xiao, D.} \& \bibinfo{author}{Cui, X.}
\newblock Valley polarization in MoS$_2$ monolayers by optical pumping.
\newblock \emph{\bibinfo{journal}{Nature Nanotechnology}} \textbf{\bibinfo{volume}{7}}, \bibinfo{pages}{490--493} (\bibinfo{year}{2012}).
\newblock \urlprefix\url{https://doi.org/10.1038/nnano.2012.95}.

\bibitem{Cao12}
\bibinfo{author}{Cao, T.} \emph{et~al.}
\newblock Valley-selective circular dichroism of monolayer molybdenum disulphide.
\newblock \emph{\bibinfo{journal}{Nature Communications}} \textbf{\bibinfo{volume}{3}}, \bibinfo{pages}{887} (\bibinfo{year}{2012}).
\newblock \urlprefix\url{https://doi.org/10.1038/ncomms1882}.

\bibitem{Rodin16}
\bibinfo{author}{Rodin, A.~S.}, \bibinfo{author}{Gomes, L.~C.}, \bibinfo{author}{Carvalho, A.} \& \bibinfo{author}{Castro~Neto, A.~H.}
\newblock Valley physics in tin (II) sulfide.
\newblock \emph{\bibinfo{journal}{Phys. Rev. B}} \textbf{\bibinfo{volume}{93}}, \bibinfo{pages}{045431} (\bibinfo{year}{2016}).
\newblock \urlprefix\url{https://link.aps.org/doi/10.1103/PhysRevB.93.045431}.

\bibitem{Lin2018}
\bibinfo{author}{Lin, S.} \emph{et~al.}
\newblock Accessing valley degree of freedom in bulk Tin(II) sulfide at room temperature.
\newblock \emph{\bibinfo{journal}{Nat. Commun.}} \textbf{\bibinfo{volume}{9}}, \bibinfo{pages}{1455} (\bibinfo{year}{2018}).
\newblock \urlprefix\url{https://doi.org/10.1038/s41467-018-03897-3}.

\bibitem{Chen18}
\bibinfo{author}{Chen, C.} \emph{et~al.}
\newblock Valley-Selective Linear Dichroism in Layered Tin Sulfide.
\newblock \emph{\bibinfo{journal}{ACS Photonics}} \textbf{\bibinfo{volume}{5}}, \bibinfo{pages}{3814--3819} (\bibinfo{year}{2018}).
\newblock \urlprefix\url{https://doi.org/10.1021/acsphotonics.8b00850}.

\bibitem{Beaulieu2024-qn}
\bibinfo{author}{Beaulieu, S.} \emph{et~al.}
\newblock Berry curvature signatures in chiroptical excitonic transitions.
\newblock \emph{\bibinfo{journal}{Science Advances}} \textbf{\bibinfo{volume}{10}}, \bibinfo{pages}{eadk3897} (\bibinfo{year}{2024}).

\bibitem{Gindl25}
\bibinfo{author}{Gindl, A.}, \bibinfo{author}{{\v{C}}mel, M.}, \bibinfo{author}{Troj{\'a}nek, F.}, \bibinfo{author}{Mal{\'y}, P.} \& \bibinfo{author}{Koz{\'a}k, M.}
\newblock Ultrafast room-temperature valley manipulation in silicon and diamond.
\newblock \emph{\bibinfo{journal}{Nat. Phys.}} \textbf{\bibinfo{volume}{21}}, \bibinfo{pages}{947--952} (\bibinfo{year}{2025}).
\newblock \urlprefix\url{https://doi.org/10.1038/s41567-025-02862-4}.

\bibitem{pan2025}
\bibinfo{author}{Pan, Y.} \emph{et~al.}
\newblock Ultrafast Strongly Anisotropic Valleytronics in SnSe.
\newblock \emph{\bibinfo{journal}{arXiv preprint}}  (\bibinfo{year}{2025}).
\newblock \eprint{2512.15400}.

\bibitem{Sharma23}
\bibinfo{author}{Sharma, S.}, \bibinfo{author}{Elliott, P.} \& \bibinfo{author}{Shallcross, S.}
\newblock THz induced giant spin and valley currents.
\newblock \emph{\bibinfo{journal}{Science Advances}} \textbf{\bibinfo{volume}{9}}, \bibinfo{pages}{eadf3673} (\bibinfo{year}{2023}).

\bibitem{Tyulnev24}
\bibinfo{author}{Tyulnev, I.} \emph{et~al.}
\newblock Valleytronics in bulk MoS$_2$ with a topologic optical field.
\newblock \emph{\bibinfo{journal}{Nature}} \textbf{\bibinfo{volume}{628}}, \bibinfo{pages}{746--751} (\bibinfo{year}{2024}).
\newblock \urlprefix\url{https://doi.org/10.1038/s41586-024-07156-y}.

\bibitem{gill2025}
\bibinfo{author}{Gill, D.}, \bibinfo{author}{Sharma, S.} \& \bibinfo{author}{Shallcross, S.}
\newblock Generation of pure, spin polarized, and unpolarized charge currents at the few cycle limit of circularly polarized light.
\newblock \emph{\bibinfo{journal}{arXiv preprint}}  (\bibinfo{year}{2025}).
\newblock \eprint{2509.18432}.

\bibitem{Zhu2025}
\bibinfo{author}{Zhu, X.} \emph{et~al.}
\newblock A holistic view of the dynamics of long-lived valley polarized dark excitonic states in monolayer WS$_2$.
\newblock \emph{\bibinfo{journal}{Nature Communications}} \textbf{\bibinfo{volume}{16}}, \bibinfo{pages}{6385} (\bibinfo{year}{2025}).
\newblock \urlprefix\url{https://doi.org/10.1038/s41467-025-61677-2}.

\bibitem{Berghauser2018}
\bibinfo{author}{Bergh{\"a}user, G.} \emph{et~al.}
\newblock Inverted valley polarization in optically excited transition metal dichalcogenides.
\newblock \emph{\bibinfo{journal}{Nature Communications}} \textbf{\bibinfo{volume}{9}}, \bibinfo{pages}{971} (\bibinfo{year}{2018}).
\newblock \urlprefix\url{https://doi.org/10.1038/s41467-018-03354-1}.

\bibitem{Lan24}
\bibinfo{author}{Lan, K.}, \bibinfo{author}{Xie, S.} \& \bibinfo{author}{Fu, J.}
\newblock Laser-field detuning assisted optimization of valley dynamics in monolayer ${\mathrm{WSe}}_{2}$.
\newblock \emph{\bibinfo{journal}{Phys. Rev. B}} \textbf{\bibinfo{volume}{110}}, \bibinfo{pages}{125420} (\bibinfo{year}{2024}).
\newblock \urlprefix\url{https://link.aps.org/doi/10.1103/PhysRevB.110.125420}.

\bibitem{Kumar2025}
\bibinfo{author}{Kumar, A.~M.} \emph{et~al.}
\newblock Strain Control of Valley Polarization Dynamics in a 2D Semiconductor via Exciton Hybridization.
\newblock \emph{\bibinfo{journal}{Nano Letters}} \textbf{\bibinfo{volume}{25}}, \bibinfo{pages}{15164--15172} (\bibinfo{year}{2025}).
\newblock \urlprefix\url{https://doi.org/10.1021/acs.nanolett.5c02636}.

\bibitem{Dai24}
\bibinfo{author}{Dai, D.} \emph{et~al.}
\newblock Twist angle–dependent valley polarization switching in heterostructures.
\newblock \emph{\bibinfo{journal}{Science Advances}} \textbf{\bibinfo{volume}{10}}, \bibinfo{pages}{eado1281} (\bibinfo{year}{2024}).

\bibitem{Wu25}
\bibinfo{author}{Wu, Y.-C.} \emph{et~al.}
\newblock Highly Tunable Valley Polarization of Potential-Trapped Moir\'e Excitons in ${\mathrm{WSe}}_{2}/{\mathrm{WS}}_{2}$ Heterojunctions.
\newblock \emph{\bibinfo{journal}{Phys. Rev. Lett.}} \textbf{\bibinfo{volume}{134}}, \bibinfo{pages}{256402} (\bibinfo{year}{2025}).
\newblock \urlprefix\url{https://link.aps.org/doi/10.1103/PhysRevLett.134.256402}.

\bibitem{Ye14}
\bibinfo{author}{Ye, Z.} \emph{et~al.}
\newblock Probing excitonic dark states in single-layer tungsten disulphide.
\newblock \emph{\bibinfo{journal}{Nature}} \textbf{\bibinfo{volume}{513}}, \bibinfo{pages}{214--218} (\bibinfo{year}{2014}).
\newblock \urlprefix\url{https://doi.org/10.1038/nature13734}.

\bibitem{Xiao2015}
\bibinfo{author}{Xiao, J.} \emph{et~al.}
\newblock Nonlinear optical selection rule based on valley-exciton locking in monolayer ws2.
\newblock \emph{\bibinfo{journal}{Light: Science {\&} Applications}} \textbf{\bibinfo{volume}{4}}, \bibinfo{pages}{e366--e366} (\bibinfo{year}{2015}).
\newblock \urlprefix\url{https://doi.org/10.1038/lsa.2015.139}.

\bibitem{Klimmer2021}
\bibinfo{author}{Klimmer, S.} \emph{et~al.}
\newblock All-optical polarization and amplitude modulation of second-harmonic generation in atomically thin semiconductors.
\newblock \emph{\bibinfo{journal}{Nature Photonics}} \textbf{\bibinfo{volume}{15}}, \bibinfo{pages}{837--842} (\bibinfo{year}{2021}).
\newblock \urlprefix\url{https://doi.org/10.1038/s41566-021-00859-y}.

\bibitem{Herrmann2025-lm}
\bibinfo{author}{Herrmann, P.} \emph{et~al.}
\newblock Nonlinear valley selection rules and all-optical probe of broken time-reversal symmetry in monolayer WSe$_2$.
\newblock \emph{\bibinfo{journal}{Nature Photonics}} \textbf{\bibinfo{volume}{19}}, \bibinfo{pages}{300--306} (\bibinfo{year}{2025}).
\newblock \urlprefix\url{https://doi.org/10.1038/s41566-024-01591-z}.

\bibitem{Friedrich2026}
\bibinfo{author}{Friedrich, F.} \emph{et~al.}
\newblock Measurement of optically induced broken time-reversal symmetry in atomically thin crystals.
\newblock \emph{\bibinfo{journal}{Nature Photonics}} \textbf{\bibinfo{volume}{20}}, \bibinfo{pages}{186--193} (\bibinfo{year}{2026}).
\newblock \urlprefix\url{https://doi.org/10.1038/s41566-025-01801-2}.

\bibitem{tornow2026}
\bibinfo{author}{Tornow, N.} \emph{et~al.}
\newblock Nonlinear Circular Dichroism Reveals the Local Berry Curvature.
\newblock \emph{\bibinfo{journal}{arXiv preprint arXiv:2604.13729}}  (\bibinfo{year}{2026}).
\newblock \urlprefix\url{https://arxiv.org/abs/2604.13729}.

\bibitem{Lin2019}
\bibinfo{author}{Lin, K.-Q.}, \bibinfo{author}{Bange, S.} \& \bibinfo{author}{Lupton, J.~M.}
\newblock Quantum interference in second-harmonic generation from monolayer WSe$_2$.
\newblock \emph{\bibinfo{journal}{Nature Physics}} \textbf{\bibinfo{volume}{15}}, \bibinfo{pages}{242--246} (\bibinfo{year}{2019}).
\newblock \urlprefix\url{https://doi.org/10.1038/s41567-018-0384-5}.

\bibitem{Lin2021}
\bibinfo{author}{Lin, K.-Q.} \emph{et~al.}
\newblock Narrow-band high-lying excitons with negative-mass electrons in monolayer WSe$_2$.
\newblock \emph{\bibinfo{journal}{Nature Communications}} \textbf{\bibinfo{volume}{12}}, \bibinfo{pages}{5500} (\bibinfo{year}{2021}).
\newblock \urlprefix\url{https://doi.org/10.1038/s41467-021-25499-2}.

\bibitem{Lin2022}
\bibinfo{author}{Lin, K.-Q.} \emph{et~al.}
\newblock High-lying valley-polarized trions in 2D semiconductors.
\newblock \emph{\bibinfo{journal}{Nature Communications}} \textbf{\bibinfo{volume}{13}}, \bibinfo{pages}{6980} (\bibinfo{year}{2022}).
\newblock \urlprefix\url{https://doi.org/10.1038/s41467-022-33939-w}.

\bibitem{Lin2021t}
\bibinfo{author}{Lin, K.-Q.} \emph{et~al.}
\newblock Twist-angle engineering of excitonic quantum interference and optical nonlinearities in stacked 2D semiconductors.
\newblock \emph{\bibinfo{journal}{Nature Communications}} \textbf{\bibinfo{volume}{12}}, \bibinfo{pages}{1553} (\bibinfo{year}{2021}).
\newblock \urlprefix\url{https://doi.org/10.1038/s41467-021-21547-z}.

\bibitem{Manca2017}
\bibinfo{author}{Manca, M.} \emph{et~al.}
\newblock Enabling valley selective exciton scattering in monolayer WSe$_2$ through upconversion.
\newblock \emph{\bibinfo{journal}{Nature Communications}} \textbf{\bibinfo{volume}{8}}, \bibinfo{pages}{14927} (\bibinfo{year}{2017}).
\newblock \urlprefix\url{https://doi.org/10.1038/ncomms14927}.

\bibitem{Fragkos2025-ob}
\bibinfo{author}{Fragkos, S.} \emph{et~al.}
\newblock Time- and polarization-resolved extreme ultraviolet momentum microscopy.
\newblock \emph{\bibinfo{journal}{Review of Scientific Instruments}} \textbf{\bibinfo{volume}{96}}, \bibinfo{pages}{115201} (\bibinfo{year}{2025}).
\newblock \urlprefix\url{https://pubs.aip.org/rsi/article/96/11/115201/3370628/Time-and-polarization-resolved-extreme-ultraviolet}.

\bibitem{Comby22}
\bibinfo{author}{Comby, A.} \emph{et~al.}
\newblock Ultrafast polarization-tunable monochromatic extreme ultraviolet source at high-repetition-rate.
\newblock \emph{\bibinfo{journal}{Journal of Optics}} \textbf{\bibinfo{volume}{24}}, \bibinfo{pages}{084003} (\bibinfo{year}{2022}).
\newblock \urlprefix\url{https://dx.doi.org/10.1088/2040-8986/ac7a49}.

\bibitem{Medjanik17}
\bibinfo{author}{Medjanik, K.} \emph{et~al.}
\newblock Direct 3D mapping of the Fermi surface and Fermi velocity.
\newblock \emph{\bibinfo{journal}{Nature Materials}} \textbf{\bibinfo{volume}{16}}, \bibinfo{pages}{615--621} (\bibinfo{year}{2017}).
\newblock \urlprefix\url{https://doi.org/10.1038/nmat4875}.

\bibitem{tkach24}
\bibinfo{author}{Tkach, O.} \& \bibinfo{author}{Schönhense, G.}
\newblock Multimode objective lens for momentum microscopy and XPEEM: Theory.
\newblock \emph{\bibinfo{journal}{Ultramicroscopy}} \textbf{\bibinfo{volume}{276}}, \bibinfo{pages}{114167} (\bibinfo{year}{2025}).
\newblock \urlprefix\url{https://www.sciencedirect.com/science/article/pii/S030439912500066X}.

\bibitem{tkach24-2}
\bibinfo{author}{Tkach, O.} \emph{et~al.}
\newblock Multimode objective lens for momentum microscopy and x-ray photoemission electron microscopy: Experiments.
\newblock \emph{\bibinfo{journal}{Review of Scientific Instruments}} \textbf{\bibinfo{volume}{97}} (\bibinfo{year}{2026}).
\newblock \urlprefix\url{http://dx.doi.org/10.1063/5.0311293}.

\bibitem{Zhang14}
\bibinfo{author}{Zhang, X.}, \bibinfo{author}{Liu, Q.}, \bibinfo{author}{Luo, J.-W.}, \bibinfo{author}{Freeman, A.~J.} \& \bibinfo{author}{Zunger, A.}
\newblock Hidden spin polarization in inversion-symmetric bulk crystals.
\newblock \emph{\bibinfo{journal}{Nature Physics}} \textbf{\bibinfo{volume}{10}}, \bibinfo{pages}{387--393} (\bibinfo{year}{2014}).
\newblock \urlprefix\url{https://doi.org/10.1038/nphys2933}.

\bibitem{Riley14}
\bibinfo{author}{Riley, J.~M.} \emph{et~al.}
\newblock Direct observation of spin-polarized bulk bands in an inversion-symmetric semiconductor.
\newblock \emph{\bibinfo{journal}{Nature Physics}} \textbf{\bibinfo{volume}{10}}, \bibinfo{pages}{835--839} (\bibinfo{year}{2014}).
\newblock \urlprefix\url{https://doi.org/10.1038/nphys3105}.

\bibitem{Razzoli17}
\bibinfo{author}{Razzoli, E.} \emph{et~al.}
\newblock Selective Probing of Hidden Spin-Polarized States in Inversion-Symmetric Bulk ${\mathrm{MoS}}_{2}$.
\newblock \emph{\bibinfo{journal}{Phys. Rev. Lett.}} \textbf{\bibinfo{volume}{118}}, \bibinfo{pages}{086402} (\bibinfo{year}{2017}).
\newblock \urlprefix\url{https://link.aps.org/doi/10.1103/PhysRevLett.118.086402}.

\bibitem{Fanciulli23}
\bibinfo{author}{Fanciulli, M.} \emph{et~al.}
\newblock Ultrafast Hidden Spin Polarization Dynamics of Bright and Dark Excitons in $2\mathrm{H}\text{\ensuremath{-}}{\mathrm{WSe}}_{2}$.
\newblock \emph{\bibinfo{journal}{Phys. Rev. Lett.}} \textbf{\bibinfo{volume}{131}}, \bibinfo{pages}{066402} (\bibinfo{year}{2023}).
\newblock \urlprefix\url{https://link.aps.org/doi/10.1103/PhysRevLett.131.066402}.

\bibitem{Beaulieu20-2}
\bibinfo{author}{Beaulieu, S.} \emph{et~al.}
\newblock Revealing Hidden Orbital Pseudospin Texture with Time-Reversal Dichroism in Photoelectron Angular Distributions.
\newblock \emph{\bibinfo{journal}{Phys. Rev. Lett.}} \textbf{\bibinfo{volume}{125}}, \bibinfo{pages}{216404} (\bibinfo{year}{2020}).
\newblock \urlprefix\url{https://link.aps.org/doi/10.1103/PhysRevLett.125.216404}.

\bibitem{Schuler22-1}
\bibinfo{author}{Sch\"uler, M.} \emph{et~al.}
\newblock Polarization-Modulated Angle-Resolved Photoemission Spectroscopy: Toward Circular Dichroism without Circular Photons and Bloch Wave-function Reconstruction.
\newblock \emph{\bibinfo{journal}{Phys. Rev. X}} \textbf{\bibinfo{volume}{12}}, \bibinfo{pages}{011019} (\bibinfo{year}{2022}).
\newblock \urlprefix\url{https://link.aps.org/doi/10.1103/PhysRevX.12.011019}.

\bibitem{Sie15}
\bibinfo{author}{Sie, E.~J.} \emph{et~al.}
\newblock Valley-selective optical Stark effect in monolayer WS$_2$.
\newblock \emph{\bibinfo{journal}{Nature Materials}} \textbf{\bibinfo{volume}{14}}, \bibinfo{pages}{290--294} (\bibinfo{year}{2015}).
\newblock \urlprefix\url{https://doi.org/10.1038/nmat4156}.

\bibitem{Sie17}
\bibinfo{author}{Sie, E.~J.} \emph{et~al.}
\newblock Large, valley-exclusive Bloch-Siegert shift in monolayer WS$_2$.
\newblock \emph{\bibinfo{journal}{Science}} \textbf{\bibinfo{volume}{355}}, \bibinfo{pages}{1066--1069} (\bibinfo{year}{2017}).
\newblock \urlprefix\url{https://www.science.org/doi/abs/10.1126/science.aal2241}.
\newblock \eprint{https://www.science.org/doi/pdf/10.1126/science.aal2241}.

\bibitem{giannozzi_quantum_2009}
\bibinfo{author}{Giannozzi, P.} \emph{et~al.}
\newblock {QUANTUM} {ESPRESSO}: a modular and open-source software project for quantum simulations of materials.
\newblock \emph{\bibinfo{journal}{J. Phys.: Condens. Matter}} \textbf{\bibinfo{volume}{21}}, \bibinfo{pages}{395502} (\bibinfo{year}{2009}).
\newblock \urlprefix\url{http://stacks.iop.org/0953-8984/21/i=39/a=395502}.

\bibitem{Mostofi08}
\bibinfo{author}{Mostofi, A.~A.} \emph{et~al.}
\newblock wannier90: A tool for obtaining maximally-localised Wannier functions.
\newblock \emph{\bibinfo{journal}{Computer Physics Communications}} \textbf{\bibinfo{volume}{178}}, \bibinfo{pages}{685--699} (\bibinfo{year}{2008}).
\newblock \urlprefix\url{https://www.sciencedirect.com/science/article/pii/S0010465507004936}.

\bibitem{girotto25}
\bibinfo{author}{Girotto~Erhardt, N.} \emph{et~al.}
\newblock Ultrafast Nonequilibrium Enhancement of Electron-Phonon Interaction in $2{\mathrm{H}\text{\ensuremath{-}}\mathrm{MoTe}}_{2}$.
\newblock \emph{\bibinfo{journal}{Phys. Rev. Lett.}} \textbf{\bibinfo{volume}{135}}, \bibinfo{pages}{146904} (\bibinfo{year}{2025}).
\newblock \urlprefix\url{https://link.aps.org/doi/10.1103/dvlz-93t8}.

\bibitem{Zhang2022-rt}
\bibinfo{author}{Zhang, Q.} \emph{et~al.}
\newblock Prolonging valley polarization lifetime through gate-controlled exciton-to-trion conversion in monolayer molybdenum ditelluride.
\newblock \emph{\bibinfo{journal}{Nat. Commun.}} \textbf{\bibinfo{volume}{13}}, \bibinfo{pages}{4101} (\bibinfo{year}{2022}).

\bibitem{Bertoni16}
\bibinfo{author}{Bertoni, R.} \emph{et~al.}
\newblock Generation and Evolution of Spin-, Valley-, and Layer-Polarized Excited Carriers in Inversion-Symmetric ${\mathrm{WSe}}_{2}$.
\newblock \emph{\bibinfo{journal}{Phys. Rev. Lett.}} \textbf{\bibinfo{volume}{117}}, \bibinfo{pages}{277201} (\bibinfo{year}{2016}).
\newblock \urlprefix\url{https://link.aps.org/doi/10.1103/PhysRevLett.117.277201}.

\bibitem{ValleyFloquet24}
\bibinfo{author}{Fragkos, S.} \emph{et~al.}
\newblock {Floquet-Bloch} valleytronics.
\newblock \emph{\bibinfo{journal}{Nat. Commun.}} \textbf{\bibinfo{volume}{16}}, \bibinfo{pages}{5799} (\bibinfo{year}{2025}).

\bibitem{verma25}
\bibinfo{author}{Verma, N.} \& \bibinfo{author}{Queiroz, R.}
\newblock Instantaneous response and quantum geometry of insulators.
\newblock \emph{\bibinfo{journal}{Proceedings of the National Academy of Sciences}} \textbf{\bibinfo{volume}{122}}, \bibinfo{pages}{e2405837122} (\bibinfo{year}{2025}).

\bibitem{Yu2025}
\bibinfo{author}{Yu, J.} \emph{et~al.}
\newblock Quantum geometry in quantum materials.
\newblock \emph{\bibinfo{journal}{npj Quantum Materials}} \textbf{\bibinfo{volume}{10}}, \bibinfo{pages}{101} (\bibinfo{year}{2025}).
\newblock \urlprefix\url{https://doi.org/10.1038/s41535-025-00801-3}.

\bibitem{Li26}
\bibinfo{author}{Li, Y.} \& \bibinfo{author}{Liu, C.-C.}
\newblock Quantum-Metric-Based Optical Selection Rules.
\newblock \emph{\bibinfo{journal}{Phys. Rev. Lett.}} \textbf{\bibinfo{volume}{136}}, \bibinfo{pages}{046901} (\bibinfo{year}{2026}).
\newblock \urlprefix\url{https://link.aps.org/doi/10.1103/bdhy-hnd2}.

\bibitem{Ahn2022}
\bibinfo{author}{Ahn, J.}, \bibinfo{author}{Guo, G.-Y.}, \bibinfo{author}{Nagaosa, N.} \& \bibinfo{author}{Vishwanath, A.}
\newblock Riemannian geometry of resonant optical responses.
\newblock \emph{\bibinfo{journal}{Nature Physics}} \textbf{\bibinfo{volume}{18}}, \bibinfo{pages}{290--295} (\bibinfo{year}{2022}).
\newblock \urlprefix\url{https://doi.org/10.1038/s41567-021-01465-z}.

\bibitem{orenstein_topology_2021}
\bibinfo{author}{Orenstein, J.} \emph{et~al.}
\newblock Topology and {Symmetry} of {Quantum} {Materials} via {Nonlinear} {Optical} {Responses}.
\newblock \emph{\bibinfo{journal}{Annu. Rev. Condens. Matter Phys.}} \textbf{\bibinfo{volume}{12}}, \bibinfo{pages}{247--272} (\bibinfo{year}{2021}).
\newblock \urlprefix\url{https://www.annualreviews.org/doi/10.1146/annurev-conmatphys-031218-013712}.

\bibitem{morimoto_topological_2016}
\bibinfo{author}{Morimoto, T.} \& \bibinfo{author}{Nagaosa, N.}
\newblock Topological nature of nonlinear optical effects in solids.
\newblock \emph{\bibinfo{journal}{Science Advances}} \textbf{\bibinfo{volume}{2}}, \bibinfo{pages}{e1501524} (\bibinfo{year}{2016}).
\newblock \urlprefix\url{https://www.science.org/doi/10.1126/sciadv.1501524}.

\bibitem{lai_third-order_2021}
\bibinfo{author}{Lai, S.} \emph{et~al.}
\newblock Third-order nonlinear {Hall} effect induced by the {Berry}-connection polarizability tensor.
\newblock \emph{\bibinfo{journal}{Nat. Nanotechnol.}} \textbf{\bibinfo{volume}{16}}, \bibinfo{pages}{869--873} (\bibinfo{year}{2021}).
\newblock \urlprefix\url{https://www.nature.com/articles/s41565-021-00917-0}.

\bibitem{aversa_nonlinear_1995}
\bibinfo{author}{Aversa, C.} \& \bibinfo{author}{Sipe, J.~E.}
\newblock Nonlinear optical susceptibilities of semiconductors: {Results} with a length-gauge analysis.
\newblock \emph{\bibinfo{journal}{Phys. Rev. B}} \textbf{\bibinfo{volume}{52}}, \bibinfo{pages}{14636--14645} (\bibinfo{year}{1995}).
\newblock \urlprefix\url{https://link.aps.org/doi/10.1103/PhysRevB.52.14636}.

\bibitem{PhysRevB.61.5337}
\bibinfo{author}{Sipe, J.~E.} \& \bibinfo{author}{Shkrebtii, A.~I.}
\newblock Second-order optical response in semiconductors.
\newblock \emph{\bibinfo{journal}{Phys. Rev. B}} \textbf{\bibinfo{volume}{61}}, \bibinfo{pages}{5337--5352} (\bibinfo{year}{2000}).
\newblock \urlprefix\url{https://link.aps.org/doi/10.1103/PhysRevB.61.5337}.

\bibitem{PhysRevLett.115.115502}
\bibinfo{author}{Zhang, L.} \& \bibinfo{author}{Niu, Q.}
\newblock Chiral Phonons at High-Symmetry Points in Monolayer Hexagonal Lattices.
\newblock \emph{\bibinfo{journal}{Phys. Rev. Lett.}} \textbf{\bibinfo{volume}{115}}, \bibinfo{pages}{115502} (\bibinfo{year}{2015}).
\newblock \urlprefix\url{https://link.aps.org/doi/10.1103/PhysRevLett.115.115502}.

\bibitem{Cheng19}
\bibinfo{author}{Cheng, J.} \emph{et~al.}
\newblock Chiral selection rules for multi-photon processes in two-dimensional honeycomb materials.
\newblock \emph{\bibinfo{journal}{Optics Letters}} \textbf{\bibinfo{volume}{44}}, \bibinfo{pages}{2141--2144} (\bibinfo{year}{2019}).
\newblock \urlprefix\url{https://opg.optica.org/ol/abstract.cfm?URI=ol-44-9-2141}.

\bibitem{Gucci2026}
\bibinfo{author}{Gucci, F.} \emph{et~al.}
\newblock Encoding and manipulating ultrafast coherent valleytronic information with lightwaves.
\newblock \emph{\bibinfo{journal}{Nature Photonics}} \textbf{\bibinfo{volume}{20}}, \bibinfo{pages}{266--272} (\bibinfo{year}{2026}).
\newblock \urlprefix\url{https://doi.org/10.1038/s41566-025-01823-w}.

\bibitem{Uzan22}
\bibinfo{author}{Uzan-Narovlansky, A.~J.} \emph{et~al.}
\newblock Observation of light-driven band structure via multiband high-harmonic spectroscopy.
\newblock \emph{\bibinfo{journal}{Nature Photonics}} \textbf{\bibinfo{volume}{16}}, \bibinfo{pages}{428--432} (\bibinfo{year}{2022}).
\newblock \urlprefix\url{https://doi.org/10.1038/s41566-022-01010-1}.

\bibitem{Uzan24}
\bibinfo{author}{Uzan-Narovlansky, A.~J.} \emph{et~al.}
\newblock Observation of interband Berry phase in laser-driven crystals.
\newblock \emph{\bibinfo{journal}{Nature}} \textbf{\bibinfo{volume}{626}}, \bibinfo{pages}{66--71} (\bibinfo{year}{2024}).
\newblock \urlprefix\url{https://doi.org/10.1038/s41586-023-06828-5}.

\bibitem{Xian20}
\bibinfo{author}{Xian, R.~P.} \emph{et~al.}
\newblock An open-source, end-to-end workflow for multidimensional photoemission spectroscopy.
\newblock \emph{\bibinfo{journal}{Scientific Data}} \textbf{\bibinfo{volume}{7}}, \bibinfo{pages}{442} (\bibinfo{year}{2020}).
\newblock \urlprefix\url{https://doi.org/10.1038/s41597-020-00769-8}.

\bibitem{Xian19_2}
\bibinfo{author}{Xian, R.~P.}, \bibinfo{author}{Rettig, L.} \& \bibinfo{author}{Ernstorfer, R.}
\newblock Symmetry-guided nonrigid registration: The case for distortion correction in multidimensional photoemission spectroscopy.
\newblock \emph{\bibinfo{journal}{Ultramicroscopy}} \textbf{\bibinfo{volume}{202}}, \bibinfo{pages}{133 -- 139} (\bibinfo{year}{2019}).
\newblock \urlprefix\url{http://www.sciencedirect.com/science/article/pii/S0304399118303474}.

\bibitem{schuler_theory_2021-1}
\bibinfo{author}{Sch\"{u}ler, M.} \& \bibinfo{author}{Sentef, M.~A.}
\newblock Theory of subcycle time-resolved photoemission: {Application} to terahertz photodressing in graphene.
\newblock \emph{\bibinfo{journal}{Journal of Electron Spectroscopy and Related Phenomena}} \textbf{\bibinfo{volume}{253}}, \bibinfo{pages}{147121} (\bibinfo{year}{2021}).
\newblock \urlprefix\url{https://www.sciencedirect.com/science/article/pii/S0368204821000736}.

\end{thebibliography}

\providecommand{\noopsort}[1]{}\providecommand{\singleletter}[1]{#1}%

\end{document}